\documentclass[12pt,a4paper]{article}
\usepackage[a4paper]{geometry}

\usepackage{amsmath}
\usepackage{amsfonts}
\usepackage{amssymb}
\usepackage{dsfont}
\usepackage{graphicx}
\usepackage[nosort]{cite}
\usepackage{hyperref}

\newcommand{\eq}[1]{\begin{equation}
                     \begin{split} #1 \end{split}
                     \end{equation}}
\newcommand{\ul}{\underline}
\newcommand{\ov}{\overline}
\newcommand{\op}{\hspace{1pt}}

\allowdisplaybreaks[2]
\numberwithin{equation}{section}

\newcommand{\dx}[1]{\ensuremath{\text{d}X^{#1}}}
\newcommand{\DX}[1]{\ensuremath{\text{D}X^{#1}}}
\newcommand{\der}{\ensuremath{\text{d}}}

\newcommand{\mder}{\ensuremath{\mathcal{D}}}

\begin{document}

\vspace*{-1.5cm}
\begin{flushright}
  {\small
  LMU-ASC 33/18\\
  MPP-2018-115 \\
  CERN-TH-2018-128
  }
\end{flushright}

\vspace{1cm}

\begin{center}
{\LARGE
Open-string T-duality and applications to \\[5pt] 
non-geometric backgrounds
}
\end{center}

\vspace{0.4cm}

\begin{center}
  Fabrizio Cordonier-Tello$^{1}$, Dieter L\"ust$^{1,2,3}$, Erik Plauschinn$^1$
\end{center}

\vspace{0.3cm}

\begin{center} 
\textit{$^{1}$\op Arnold Sommerfeld Center for Theoretical Physics\\[1pt]
Ludwig-Maximilians-Universit\"at \\[1pt]
Theresienstra\ss e 37 \\[1pt]
80333 M\"unchen, Germany}
\\[1em]
\textit{$^{2}$\op Max-Planck-Institut f\"ur Physik\\[1pt]
F\"ohringer Ring 6 \\[1pt]
80805   M\"unchen, Germany}
\\[1em]
\textit{$^{3}$\op CERN, Theory Department \\[1pt]
1211 Geneva 23, Switzerland}
\end{center} 

\vspace{0.5cm}


\begin{abstract}
\noindent
We revisit T-duality transformations for the open string 
via Buscher's procedure 
and work-out technical details which have been missing so far in the literature. 
We take into account non-trivial topologies of the world-sheet, we consider
T-duality along directions with Neumann as well as Dirichlet boundary conditions, and we include 
collective T-duality along multiple directions.

\noindent
We illustrate this formalism with the example of the three-torus with $H$-flux and 
its T-dual backgrounds, and we discuss global properties of open-string  boundary conditions 
on such spaces.

\end{abstract}


\clearpage

\tableofcontents


\section{Introduction}

Dualities are remarkable features of string theory which have helped to expose and understand some of 
the intricate structures of the theory. Examples for dualities  are 
S-duality, T-duality, mirror symmetry and heterotic--type I duality --- and in this work 
we will be interested in T-duality for open strings.


\subsubsection*{Non-geometric backgrounds}

Dualities can be used to construct new backgrounds for string theory which are 
well-defined only using duality transformations. An example is F-theory in which  S-duality is 
needed to construct globally-defined solutions, and  
in a similar way  T-duality can be utilized  to build so-called non-geometric backgrounds.
The latter are spaces which do not allow for a description in terms of Riemannian geometry,
but in which  $O(D,D;\mathbb Z)$ T-duality transformations are used as transition functions between 
local charts \cite{Hellerman:2002ax,Hull:2004in}.
In this way globally well-defined backgrounds can be constructed.

The standard example for a non-geometric background is obtained by applying 
successive or collective T-duality transformations to a three-torus with 
$H$-flux. After one T-duality one arrives at 
a twisted three-torus \cite{Dasgupta:1999ss,Kachru:2002sk}, which is a geometric space and to which one 
can associate a geometric flux $F^k{}_{ij}$. 
A second T-duality transformation gives the T-fold background \cite{Hull:2004in}, which 
locally can be expressed in terms of a metric and $B$-field but which 
globally is non-geometric. To this configuration  one can associate a 
$Q$-flux $Q_i{}^{jk}$. Finally, even though the Buscher rules cannot be applied, 
a formal third T-duality leads to a background with  $R$-flux $R^{ijk}$ \cite{Shelton:2005cf,Shelton:2006fd}
and it has been argued that this  space not even locally allows for a geometric description. 
This chain of T-duality  transformations is often summarized as follows
\eq{
  \label{ex1_chain}
  H_{ijk} \qquad\xrightarrow{\hspace{10pt}T_i\hspace{10pt}}\qquad
  F^i{}_{jk}{} \qquad\xrightarrow{\hspace{10pt}T_j\hspace{10pt}}\qquad
  Q_{k}{}^{ij} \qquad\xrightarrow{\hspace{10pt}T_k\hspace{10pt}}\qquad
  R^{ijk} \,,
}
where $T_i$ denotes a T-duality transformation along the direction labelled by $i$. 
Non-geometric backgrounds have interesting properties, for instance, 
they lead to non-commutative 
\cite{
Mathai:2004qq,
Mathai:2004qc,
Lust:2010iy,
Condeescu:2012sp,
Andriot:2012an,
Andriot:2012vb,
Blair:2014kla
}
and non-associative  
\cite{
Bouwknegt:2004ap,
Ellwood:2006my,
Blumenhagen:2010hj,
Blumenhagen:2011ph,
Plauschinn:2012kd,
Mylonas:2012pg,
Bakas:2013jwa,
Mylonas:2013jha,
Chatzistavrakidis:2015vka
}
structures (for a review  see \cite{Szabo:2018hhh}).
Furthermore, non-geometric $Q$- and $R$-fluxes can help  in stabilizing moduli
in string-phenomenology 
(see e.g. \cite{
Shelton:2005cf,
Aldazabal:2006up,
Villadoro:2006ia,
Shelton:2006fd,
Micu:2007rd,
Font:2008vd,
Caviezel:2009tu,
Dibitetto:2011gm
})
and they can provide mechanisms for the  construction of potentials for inflationary scenarios in string-cosmology
(see e.g. \cite{
Hassler:2014mla,
Blumenhagen:2015kja
}).

The purpose of this paper is to investigate non-geometric backgrounds from
an open-string perspective. In particular, we want to understand the global properties of 
D-branes in non-geometric spaces, which is important for 
applications in D-brane model building. 
In the context of doubled geometry, D-branes in non-geometric backgrounds 
have been discussed already  in 
the original paper \cite{Hull:2004in} and have been investigated further in 
\cite{Lawrence:2006ma,Albertsson:2008gq}.
However, here we are interested in a description in string theory.


\subsubsection*{T-duality for open strings}

In order to analyze D-branes in non-geometric backgrounds, we 
follow a strategy similar to the closed-string situation and apply 
T-duality transformations to D-brane configurations on a three-torus with $H$-flux.

T-duality for open strings in toroidal backgrounds with constant metric and 
$B$-field can be studied via well-known conformal-field theory (CFT) techniques, 
but for curved backgrounds with non-trivial $B$-field one usually has to 
rely on Buscher's procedure \cite{Buscher:1987sk} of gauging 
a symmetry of the world-sheet action and integrating-out the gauge field. 
This approach has been investigated already in the literature, 
and some of the relevant papers are the following:
\begin{itemize}

\item In \cite{Alvarez:1996up} the authors discuss T-duality for the open string along 
a single direction. They consider the bosonic string as well as the superstring 
from a world-sheet point of view, and they analyze the behavior of the 
effective target-space action.
The direction along which the T-duality transformation is performed has Neumann 
boundary conditions, and a Lagrange multiplier is implemented in a way which does 
not allow for a generalization to Dirichlet directions for non-trivial world-sheet topologies.

\item In \cite{Dorn:1996an,Dorn:1997if} the case of a single T-duality transformation is studied from 
a path-integral point of view and as a canonical transformation. 
The gauge group on a stack of D-branes can be non-abelian, 
and the $B$-field is required to be independent of the coordinate along which T-duality is 
performed. T-duality is considered along a 
direction with Neumann boundary condition, and the topology of the world-sheet is assumed to be trivial.

\item In \cite{Forste:1996hy,Forste:1996ai,Forste:1996ms} the authors investigate T-duality along multiple directions with a
non-abelian isometry group. The $B$-field and consequently the $H$-flux are set to zero, 
and the topology of the world-sheet
is  assumed to be trivial.

\item T-duality for open-string sigma models has been studied also in \cite{Albertsson:2004gr} 
with the inclusion of the fermionic sector on the world-sheet. Boundary conditions are 
discussed and the case of a single T-duality is considered for world-sheets with trivial topology.

\end{itemize}
In the present paper we extend the above analyses and work-out  missing details: 1) we 
include world-sheets with a non-trivial topology in our studies, 2) we present 
procedures for T-duality transformations along directions with Neumann as well as Dirichlet boundary conditions, and 3) 
we include the case of performing collective T-duality transformations along multiple directions. 
For part of our analysis we include the case of a non-abelian isometry algebra, 
but final results are obtained only for the abelian case.

We furthermore mention that T-duality for open strings has been studied 
via canonical transformations also in
\cite{Borlaf:1996na,Lozano:1996sc}, and 
T-duality for D-branes from an effective field theory point of view has been 
investigated for instance in \cite{Tseytlin:1996it,Bergshoeff:1996cy,Green:1996bh}.
D-branes in non-geometric backgrounds have been analyzed from an effective field theory point of view
in \cite{Ellwood:2006my} and through boundary states in \cite{Kawai:2007qd},
and in the context of generalized complex geometry D-branes and T-duality have been studied
in \cite{Grange:2005nm}.
Poisson-Lie T-duality for open strings has been discussed in \cite{Klimcik:1995np}
and \cite{Albertsson:2006zg},  and in a somewhat different approach 
T-duality for open strings has also been considered in 
\cite{Davidovic:2016xhh,Sazdovic:2016ley,Sazdovic:2017lqo}.


\subsubsection*{Outline}

This paper consists of essentially two parts: first we discuss on general grounds 
open-string T-duality via Buscher's procedure, and in the second part we apply this formalism 
to D-brane configurations on a  three-torus with $H$-flux.
More concretely:
\begin{itemize}

\item In section~\ref{sec:1} we consider the world-sheet description of 
open strings, the gauging of symmetries and how the ungauged action 
can be obtained from the gauged one. We pay special attention to 
the topology of the world-sheet. 

\item In section~\ref{sec_t-duality} we perform collective T-duality transformations
either along all Neumann or all Dirichlet directions. 
We find for instance that, as expected, under T-duality Neumann and Dirichlet boundary conditions 
are interchanged.

\item In section~\ref{sec_examples} we consider D-branes on a three-torus with $H$-flux 
and study T-duality along one, two and three directions. 
This is on the one-hand to illustrate the procedure discussed in section~\ref{sec_t-duality}, 
and on the other-hand to obtain explicit examples for D-branes on the twisted three-torus and
on the T-fold.

\item In section~\ref{sec_discussion} we discuss the results from section~\ref{sec_examples}.
We review the Freed-Witten anomaly cancellation condition, and we show that D-branes 
satisfying this condition are globally well-defined on the twisted torus and on the T-fold.

\item Section~\ref{sec_sum_con} contains a brief summary of the results obtained in 
this paper.

\end{itemize}


\clearpage
\section{Non-linear sigma-model for the open string}\label{sec:1}

We begin our discussion by reviewing the non-linear sigma-model description of the 
open string and studying the gauging of world-sheet symmetries.  The latter are 
then employed in  section~\ref{sec_t-duality} for performing T-duality transformations.


\subsection{World-sheet action}

We first introduce the sigma-model for the open string, specify boundary conditions and 
consider global symmetries of the world-sheet action. 


\subsubsection*{The action}

The world-sheet action for an open string can be defined on a two-dimensional world-sheet 
$\Sigma$ with non-empty boundary $\partial\Sigma\neq\varnothing$. 
We allow for a non-trivial target-space metric $G$,  Kalb-Ramond field $B$, 
dilaton $\phi$ and 
open-string gauge field $a$, although at a later stage we
impose restrictions upon them.  The field strengths of $B$ and $a$ will 
be denoted by $H=\der B$ and $F=\der a$, respectively.
For later convenience we  perform a Wick rotation and consider 
an Euclidean metric on the world-sheet. 
The resulting action then reads
\eq{
\label{0}
\arraycolsep2pt
\begin{array}{lcll}
 \mathcal S&=& \displaystyle-\frac{1}{2\pi\alpha'}\int_{\Sigma} & 
 \displaystyle \biggl[ \, 
 \frac{1}{2}\op G_{ij}\op \dx{i}\wedge\op\star\dx{j}+
 \frac{i}{2}\op B_{ij}\op \dx{i}\wedge\dx{j}
 +\frac{\alpha'}{2} \mathsf R\, \phi\star 1 
 \,\biggr]  
 \\[12pt]
 &&\displaystyle-\frac{1}{2\pi\alpha'}\int_{\partial\Sigma} &\biggl[\, 2\pi i\op \alpha'\op a_i \op \dx{i}+\alpha' \op k(s) \op
   \phi \, \der s
 \,\biggr] \,,
 \end{array}
}
where  $a= a_i\op \dx{i}$ is understood to be restricted to the boundary $\partial\Sigma$.
Coordinates on the world-sheet $\Sigma$ are denoted as $\sigma^{\mathsf a} = \{\sigma^1,\sigma^2\}$
and on the boundary $\partial\Sigma$ as $s$. 
Coordinates on a $D$-dimensional target-space are denoted by $X^i(\sigma)$ with $i=1,\ldots,D$,
which can be interpreted as maps from the world-sheet $\Sigma$ to the target-space.
The exterior derivative operator $\der $ acting on $X^i$ can therefore be written as 
$\der X^i=\partial_{\mathsf a} X^i(\sigma)\, \der \sigma^{\mathsf a}$.
The Hodge star-operator on $\Sigma$ is denoted by $\star$, 
the Ricci scalar for the world-sheet metric is $\mathsf R$,
and $k(s)$ is the extrinsic 
curvature of the boundary $k=t^{\mathsf a}\op t^{\mathsf b}\op\nabla_{\mathsf a} n_{\mathsf b}$, 
where $t^{\mathsf a}$ 
and $n^{\mathsf a}$ are unit vectors tangential and normal to the boundary, respectively. 
Note furthermore that on the boundary we have $\dx{i} \rvert_{\partial\Sigma}= t^{\mathsf a}\partial_{\mathsf a} X^i \op \der s$.


\subsubsection*{Boundary conditions}

Next, we consider the equations of motion for the fields $X^i$ and, in particular, the 
corresponding boundary conditions. Denoting by $\Gamma^i_{jk}$ the Christoffel symbols
for the target-space metric $G_{ij}$  and by $H_{ijk}$ the components of the field strength $H=\der B$, the equations 
of motion for $X^i$ following from \eqref{0} read
\eq{
  \label{eom_x_001}
  0 = \der\star \dx{i} + \Gamma^i_{mn} \op\dx{m}\wedge \star \dx{n}
  &- \frac{i}{2} \op H^{i}{}_{mn}\op \dx{m}\wedge \dx{n} \\
  &- \frac{\alpha'}{2}\op G^{im} \op\partial_m \phi \, \mathsf R \star 1\,,
}
where the index of $H_{ijk}$ has been raised using the inverse of $G_{ij}$. 
As boundary condition we can impose 
Dirichlet boundary conditions of the form 
$\delta X^{i} \rvert_{\partial\Sigma}=0$
or Neumann boundary conditions. 
Denoting the tangential and normal part of $\dx{i}$ on the boundary by
\eq{
  \bigl(\dx{i}\bigr)_{\rm tan}  \equiv
   t^{\mathsf a} \op\partial_{\mathsf a} X^i \, \der s \bigr\rvert_{\partial\Sigma} \,,
   \hspace{50pt}
  \bigl(\dx{i}\bigr)_{\rm norm}  \equiv
   n^{\mathsf a} \op\partial_{\mathsf a} X^i \, \der s \bigr\rvert_{\partial\Sigma} \,,
}
where $t^{\mathsf a}$ and $n^{\mathsf a}$ are again unit tangential and normal vectors, 
and introducing the gauge-invariant open-string field-strength 
$\mathcal F$ as
$2\pi\alpha'  \mathcal F = 2\pi \alpha' F + B$,
we can summarize the boundary conditions as
\eq{
  \label{boundary_cond}
  \arraycolsep2pt
  \begin{array}{@{}l@{\hspace{30pt}}l@{}}
  \mbox{Dirichlet} & \displaystyle 0 = \bigl(\dx{\hat i}\bigr)_{\rm tan}  \,,
  \\[6pt]
  \mbox{Neumann} & \displaystyle 0 = 
   G_{ai} \op\bigl(\dx{i}\bigr)_{\rm norm} + 2\pi \alpha' i \op \mathcal F_{ab} \bigl(\dx{b}\bigr)_{\rm tan}
  + \alpha'\op k(s) \op  \partial_a \phi\,\der s\op\Bigr\rvert_{\partial\Sigma} \,.
  \end{array}
}
Here and in the following we split the target-space index $i=1,\ldots, D$ into Dirichlet and Neumann directions 
$\hat i $ and $a$, respectively.

Now, the Hodge decomposition theorem for manifolds with boundary 
(see for instance \cite{Cappell2006}) allows us to decompose the space of 
closed $p$-forms $C^p= \{\omega \in\Omega^p: \der\omega=0\}$ into exact 
$p$-forms $E^p= \{\omega \in\Omega^p: \omega = \der\eta, \op \eta\in \Omega^{p-1}\}$ 
and closed and co-closed forms whose normal part vanishes on the boundary
$CcC^p_N= \{\omega\in\Omega^p: \der\omega=0,\der^{\dagger} \omega = 0, \omega_{\rm norm}=0\}$.
Here, $\Omega^p$ is the space of smooth differential forms on $\Sigma$ and
$\der^{\dagger}$ denotes the co-differential. In formulas, this decomposition reads
\eq{
 \label{hodge_decomp}
 C^p = E^p \oplus CcC^p_N \,.
}
For Dirichlet boundary conditions -- with vanishing tangential part of $\dx{\hat i}$ -- this implies in particular
that the $CcC^1_N$ part of $\dx{\hat i}$ vanishes and hence $\dx{\hat i}$ is an exact one-form on the 
world-sheet $\Sigma$.


\subsubsection*{Global symmetries}

In order to perform T-duality transformations, 
let us now require the world-sheet action \eqref{0} to be invariant under 
global symmetry transformations of the form
\eq{
  \label{1}
  \delta_{\epsilon} X^i = \epsilon^{\alpha}\op k_{\alpha}^i  \,.
}
For global symmetries the infinitesimal transformation parameters $\epsilon^{\alpha}$ are constant,
and we require the target-space vectors 
$k_{\alpha}$ to satisfy a Lie algebra $\mathfrak g$ as
\begin{equation}
  \label{2}
 [k_\alpha,k_\beta]_{L}=f_{\alpha\beta}{}^\gamma \op k_\gamma \,,
\end{equation}
where $f_{\alpha\beta}{}^{\gamma}$ are the structure constants
satisfying the Jacobi identity. The label $\alpha$ takes values $\alpha = 1,\ldots, N$ with $N=\dim(\mathfrak g)$.

The variation of the action \eqref{0} with respect to the transformations \eqref{1}  does 
not vanish in general. However, when the following conditions are met then \eqref{1}  is 
a global symmetry of the action
\eq{
  \label{constr1}
  \arraycolsep2pt
  \begin{array}{lcl@{\hspace{60pt}}lcl}
 \mathcal{L}_{k_\alpha}G&=& 0 \,,  \\[6pt]
 \mathcal{L}_{k_\alpha}B&=&\der v_\alpha \,, 
 & \displaystyle  2\pi\alpha'\mathcal{L}_{k_\alpha}a\op \bigr\rvert_{\partial\Sigma} &=&\displaystyle 
 \left(-v_\alpha+\der \omega_\alpha \right) \bigr\rvert_{\partial\Sigma} \,,
 \\[6pt]
 \mathcal{L}_{k_\alpha}\phi&=& 0 \,.
   \end{array}
}
Here, $G = \tfrac{1}{2} \op G_{ij}\op \dx{i}\otimes \dx{j}$, $B=\tfrac{1}{2} \op B_{ij} \op\dx{i}\wedge \dx{j}$,
$a = a_i \op\dx{i}$ and $\phi$ are interpreted as target-space quantities.\footnote{In order to keep our
notation manageable, we do not explicitly distinguish between quantities on the world-sheet and on target space,
but assume this to be clear from the context.
For instance, on the world-sheet we have $\mathcal L_k G = \frac{1}{2}\op (\mathcal L_k G)_{ij} 
dX^i\wedge\star dX^j$. 
}
The Lie derivative along a vector field $k$
is given by $\mathcal{L}_{k}=\der \circ\iota_k+\iota_k\circ\der$,
with $\iota$ the contraction operator acting on $\dx{i}$ as $\iota_{\partial_i} \dx{j} = \delta_i{}^j$
and $\der$ is the exterior derivative acting as $\der \equiv \dx{i} \, \partial_i$. 
Furthermore, in order to apply Stokes' theorem and show that \eqref{1} is a global symmetry 
of the action \eqref{0}
we require 
\eq{
  \begin{array}{lcll}
  v_{\alpha} &\ldots& \mbox{globally-defined one-forms} & \mbox{on~}\Sigma\,,
  \\[4pt]
  \omega_{\alpha} &\ldots& \mbox{globally-defined functions} & \mbox{on~}\partial\Sigma\,.
  \end{array}
}

Let us also investigate when the  boundary conditions \eqref{boundary_cond} are invariant 
under the global symmetry \eqref{1}. 
For Neumann boundary conditions  there are no restrictions due to \eqref{constr1},
whereas for Dirichlet boundary conditions  we find a non-trivial requirement. These 
are summarized as 
\eq{
  \label{bc_global}
  \arraycolsep2pt
  \begin{array}{@{}l@{\hspace{40pt}}l@{}}
  \mbox{Dirichlet} & \displaystyle 0 = \partial_a k_{\alpha}^{\hat i}\op \Bigr\rvert_{\partial\Sigma} \, ,
  \\[6pt]
  \mbox{Neumann} & \displaystyle \varnothing \,.  
  \end{array}
}
However, strictly speaking the Dirichlet conditions read  $\delta X^{\hat i} \rvert_{\partial\Sigma}=0$
which are not preserved under global transformations of the form \eqref{1}. 
For local symmetry transformations on the other hand, the situation is different and we can preserve 
 Dirichlet conditions of this form.


\subsection{Gauged world-sheet action}
\label{sec_gauging_ws}

In order to follow Buscher's approach to T-duality, we now  promote the global symmetries \eqref{1} to local ones by 
introducing corresponding gauge fields. 


\subsubsection*{Gauged action I}

To gauge the global symmetries, we consider $\epsilon^{\alpha} \equiv \epsilon^{\alpha}(\sigma^{\mathsf a})$ and introduce
world-sheet gauge fields $A^{\alpha}$.
The resulting gauged action takes the following form
\eq{
\label{10}
\arraycolsep2pt
\begin{array}{@{\hspace{-10pt}}lcll@{\hspace{2pt}}l@{\hspace{-10pt}}}
 \widehat{\mathcal S}&=&\displaystyle-\frac{1}{2\pi\alpha'}\int_{\Sigma}& 
 \displaystyle \biggl[ \, \frac{1}{2}\op G_{ij}\op
 \bigl(\dx{i}+k^i_{\alpha}A^\alpha\bigr)\wedge\star\bigl(\dx{j}+k^j_{\beta}A^\beta\bigr)
 +\frac{\alpha'}{2} \mathsf R\, \phi\star 1 &\biggr]  
 \\[14pt]
 &&\displaystyle-\frac{i}{2\pi\alpha'}\int_{\Sigma}&
 \displaystyle \biggl[ \, \frac{1}{2}\op B_{ij}\,
 \dx{i}\wedge\dx{j}  
 \\
 &&&\hspace{13.5pt} \displaystyle + (\tilde{v}_\alpha
 +\der \chi_\alpha)\wedge A^\alpha+\frac{1}{2}
 \left(\iota_{k_{[\underline{\alpha}}}\tilde{v}_{\underline{\beta}]}+
 f_{\alpha\beta}{}^{\gamma}\chi_{\gamma}\right)A^\alpha\wedge A^\beta &\biggr] 
 \\[14pt]
 &&\displaystyle-\frac{1}{2\pi\alpha'}\int_{\partial\Sigma} & 
 \displaystyle \biggl[ \,2\pi i\op \alpha'\op a_a\op
 \dx{a}-i \,\Omega_{\partial\Sigma} + \alpha' \op k(s) \op
   \phi \, \der s &\biggr]\,,
\end{array}
}
where  for later purpose we introduced a set of scalar fields $\chi_{\alpha}$ 
with $\alpha = 1,\ldots, N$  and where we have defined 
\eq{
  \label{def_v_002}
  \tilde{v}_\alpha:=v_{\alpha}-\imath_{k_\alpha}B \,.
}  
The one-form $\Omega_{\partial\Sigma}$ depends on whether the local symmetries are along 
Neumann or Dirichlet directions and we specify its precise form below. 
The local symmetry transformations for the gauged action \eqref{10} are given as follows\footnote{Our 
convention for the (anti-)sym\-me\-tri\-zation of indices contains a factor of $1/n!$, and 
for better distinction we highlight the (anti-)symmetrized indices by  under-lining or over-lining them.}
\eq{
  \label{15}
  \arraycolsep2pt
  \begin{array}{lclcl}
  \hat{\delta}_\epsilon X^i &=& \displaystyle \epsilon^{\alpha}\op k^i_{\alpha} \,, \\[4pt]
  \hat \delta_\epsilon A^\alpha&=&\displaystyle  -\der \epsilon^\alpha
 &-& \displaystyle  f_{\beta\gamma}{}^{\alpha}\op \epsilon^{\beta}\op A^{\gamma} \,, \\[4pt]
 \hat \delta_\epsilon \chi_\alpha &=&
 \displaystyle  -\iota_{k_{(\ov{\alpha}}}v_{\ov{\beta})}\op \epsilon^\beta
 &-&\displaystyle f_{\alpha\beta}{}^{\gamma}\op \epsilon^\beta \op\chi_\gamma\,, 
 \end{array}
}
however, under \eqref{15} the gauged action \eqref{10} is in general not invariant. 
For invariance of the terms in the bulk $\Sigma$
we have to require that  $\der\chi_{\alpha}$ are globally well-defined on $\Sigma$
and have to impose the additional constraints 
\eq{
\label{16_1}
\mathcal L_{k_{[\ul \alpha}} \tilde v_{\ul \beta]} = f_{\alpha\beta}{}^{\gamma} \tilde v_{\gamma} \,,
\hspace{40pt}
\iota_{k_{[\ul \alpha}} \op f_{\ul \beta\ul \gamma ]}{}^{\delta} \tilde v_{\delta} = \frac{1}{3} \,
\iota_{k_{\alpha}}\iota_{k_{\beta}}\iota_{k_{\gamma}} H \,.
}


\subsubsection*{Boundary conditions}

Let us now come to the  boundary conditions for the gauged action. 
In particular, the conditions \eqref{boundary_cond} 
for the fields $X^i$
are preserved provided that the transformation parameters $\epsilon^{\alpha}$ satisfy on 
the boundary
\eq{
  \label{bc_local_1}
  \arraycolsep2pt
  \begin{array}{@{}l@{\hspace{40pt}}l@{}}
  \mbox{Dirichlet} & \displaystyle 0 = k_{\alpha}^{\hat i} \bigl( \der\epsilon^{\alpha}\bigr)_{\rm tan} 
  \op\Bigr\rvert_{\partial\Sigma}\,,
  \\[10pt]
  \mbox{Neumann} & \displaystyle 0 = 
  G_{ai}\op k^i_{\alpha} \op\bigl(\der \epsilon^{\alpha}\bigr)_{\rm norm} + 2\pi \alpha' i \op \mathcal F_{ab} \op 
  k_{\alpha}^b
  \bigl(\der \epsilon^{\alpha}\bigr)_{\rm tan} \op\Bigr\rvert_{\partial\Sigma}\,,
  \end{array}
}
where we employed the  restrictions \eqref{bc_global} coming from the global symmetries.
However, for Dirichlet boundary conditions we again  have a stronger requirement from demanding 
that $\delta X^{\hat i} \rvert_{\partial\Sigma}=0$. This implies in particular
that $k_{\alpha}^{\hat i} \epsilon^{\alpha} \rvert_{\partial\Sigma}=0$.

We now turn to the boundary conditions for the gauge fields $A^{\alpha}$ and start with the following two observations:
\begin{itemize}

\item For Dirichlet conditions we argued that the corresponding transformation parameters $\epsilon^{\alpha}$ 
have to vanish on the boundary. Comparing with the transformations \eqref{15}, we can conclude that under 
local symmetry transformations the gauge fields $A^{\alpha}$ do not change on the boundary. 
In fact, as we will discuss in section~\ref{sec_recovery},
in order to show the equivalence to the ungauged action we  demand that $A^{\alpha}$ 
vanishes on the boundary.

\item For Neumann conditions we can determine the equations of motion for $X^i$ from the gauged action 
\eqref{10}. If we require that for this variation boundary terms vanish, we obtain 
conditions for  the $A^{\alpha}$ summarized below.

\end{itemize}
Motivated by these observations, we then impose the following boundary conditions for the 
gauge fields $A^{\alpha}$ \cite{Albertsson:2004gr} 
\eq{
  \label{bc_local_2}
  \arraycolsep2pt
  \begin{array}{@{}l@{\hspace{40pt}}l@{}}
  \mbox{Dirichlet} & \displaystyle 0 = k_{\alpha}^{\hat i} \bigl(  A^{\alpha}\bigr)_{\rm tan} 
  \op\Bigr\rvert_{\partial\Sigma}\,,
  \\[8pt]
  \mbox{Neumann} & \displaystyle 0 = 
  G_{ai}\op k^i_{\alpha} \op\bigl(A^{\alpha}\bigr)_{\rm norm} + 2\pi \alpha' i \op \mathcal F_{ab} \op 
  k_{\alpha}^b
  \bigl(A^{\alpha}\bigr)_{\rm tan} \op\Bigr\rvert_{\partial\Sigma}\,.
  \end{array}
}
Note that these conditions are preserved under the local symmetry transformations \eqref{15}.


\subsubsection*{Gauged action II}

After having discussed the boundary conditions for the gauge fields, we can now 
specify the one-form $\Omega_{\partial\Sigma}$ in the  
action \eqref{10}. For simplicity we consider local symmetry transformations either all along 
Dirichlet directions $\raisebox{0pt}[0pt][0pt]{$X^{\hat i}$}$ or all along Neumann directions $X^a$. 
Mixed cases can also be treated, but the presentation becomes more involved. 
\begin{itemize}

\item We start with  Dirichlet boundary conditions. In this case the infinitesimal variation parameters 
$\epsilon^{\alpha}$ vanish on the boundary, and hence the boundary terms in \eqref{10}
stay invariant under \eqref{15}.
For $\Omega_{\partial\Sigma}$ we then choose
\eq{
  \label{dirc_7347375}
  \arraycolsep2pt
  \begin{array}{@{}l@{\hspace{45pt}}l@{\hspace{149pt}}}
  \mbox{Dirichlet} & \displaystyle  \Omega_{\partial\Sigma} =0\,.
  \end{array}
}

\item For Neumann boundary conditions we introduce a second set of Lagrange 
multipliers $\phi_{\alpha}$ with $\alpha = 1,\ldots, N$, and we specify the 
one-form $\Omega_{\partial\Sigma}$ as
\eq{
  \label{neum_2846}
  \arraycolsep2pt
  \begin{array}{@{}l@{\hspace{40pt}}l@{}}
  \mbox{Neumann} & \displaystyle  \Omega_{\partial\Sigma} =
  \bigl(\chi_\alpha+\phi_{\alpha} + \omega_\alpha - 2\pi \op \alpha' \iota_{k_{\alpha}} a\bigr)\op A^\alpha \,.
  \end{array}
}
The $\phi_{\alpha}$ are required to be constant fields on the boundary $\partial\Sigma$,
and the  $\chi_\alpha$ have to be globally well-defined 
on the boundary $\partial\Sigma$.
The latter requirement implies that $\der\chi_{\alpha}$ is exact on the boundary, and
hence it follows from the Hodge decomposition  \eqref{hodge_decomp}
that $CcC_N$-part of $\der\chi_{\alpha}$ vanishes. 
We therefore have in summary 
\eq{
  \label{hodge_004}
  \begin{array}{lclrlr}
  \chi_{\alpha} &\ldots& \mbox{globally-defined} & \mbox{functions} & \mbox{on~}&\Sigma\,, \\[4pt]
  \phi_{\alpha} &\ldots& \mbox{constant} & \mbox{functions} & \mbox{on~}&\partial\Sigma\,.
  \end{array}
}
Finally, in order for the gauged action \eqref{10} to be invariant under the symmetry 
transformations \eqref{15} in the case of Neumann boundary conditions, in addition to \eqref{16_1} we impose
\eq{
\label{16_2}
\mathcal{L}_{k_{[\underline{\alpha}}}\omega_{\underline{\beta}]}  \Bigr\rvert_{\partial\Sigma} =
\frac{1}{2} \,\Bigl[  f_{\alpha\beta}{}^\gamma\omega_\gamma + 
 \iota_{k_{[\ul \alpha}} v_{\ul \beta]} \Bigr]
\, \Bigr\rvert_{\partial\Sigma}\,,
\hspace{40pt}
0=f_{\alpha\beta}{}^\gamma\phi_\gamma\,\Bigr\rvert_{\partial\Sigma}\,.
}

\end{itemize}


\subsubsection*{Symmetries of the gauged action}

The gauged action \eqref{10} has been constructed such that it is invariant under the 
local transformations \eqref{1}. However, by extending the original action by
additional fields $v_{\alpha}$, $\omega_{\alpha}$ and $\phi_{\alpha}$, further symmetries may arise. 
And indeed, we find the following transformations which leave the action \eqref{10} invariant \cite{Forste:1996hy}:
\begin{enumerate}

\item Gauge transformations of the Kalb-Ramond field 
with a globally well-defined one-form on the world-sheet $\Sigma$ denoted by $\Lambda$:
\eq{
  \arraycolsep2pt
 \begin{array}{lcl}
 \displaystyle B&\rightarrow&  \displaystyle B+\der\Lambda\,, 
 \\[4pt]
 \displaystyle  a&\rightarrow&  \displaystyle a-\tfrac{1}{2\pi\alpha'}\op \Lambda \,, 
 \\[4pt]
 \displaystyle  v_\alpha&\rightarrow&  \displaystyle v_\alpha+\iota_{k_{\alpha}}\der\Lambda\,, 
 \\[4pt]
 \displaystyle  \omega_\alpha&\rightarrow&  \displaystyle \omega_\alpha-\iota_{k_{\alpha}}\Lambda \,.
 \end{array}
}

\item Shifts of the one-forms $v_{\alpha}$ by exact forms using functions $\lambda_{\alpha}$:
\eq{
  \arraycolsep2pt
 \begin{array}{lcl}
 \displaystyle  v_\alpha&\rightarrow& \displaystyle v_\alpha+\der\lambda_\alpha \,,
 \\[4pt]
 \displaystyle \chi_\alpha&\rightarrow& \displaystyle \chi_\alpha-\lambda_\alpha \,, 
 \\[4pt]
 \displaystyle \omega_\alpha&\rightarrow& \displaystyle \omega_\alpha+\lambda_\alpha \,,
 \end{array}
 \hspace{60pt}
 \mathcal L_{k_{[\ul \alpha}} \lambda_{\ul\beta]} = f_{\alpha\beta}{}^{\gamma} \lambda_{\gamma} \,.
}

\item Gauge transformations of the open-string gauge field $a$ 
with a globally well-defined function $\lambda$  on the boundary $\partial\Sigma$:
\eq{
  \arraycolsep2pt
 \begin{array}{lcl}
 \displaystyle  a&\rightarrow& \displaystyle a+\der\lambda \,, 
 \\[4pt]
 \displaystyle\omega_\alpha&\rightarrow&  \displaystyle\omega_\alpha+2\pi\alpha'\imath_{k_\alpha}\der\lambda \,.
 \end{array}
}

\item Shifts of the functions $\omega_{\alpha}$ by constants $\theta_{\alpha}$:
\eq{
  \arraycolsep2pt
 \begin{array}{lcl}
 \displaystyle   \chi_\alpha&\rightarrow&\displaystyle  \chi_\alpha+\theta_\alpha \,,
 \\[4pt]
 \displaystyle  \omega_\alpha&\rightarrow& \displaystyle   \omega_\alpha-\theta_\alpha \,,
  \end{array}
  \hspace{60pt}
    f_{\alpha\beta}{}^{\gamma} \op \theta_{\gamma} = 0 \,.
}

\item Shifts of the functions $\phi_{\alpha}$ by constants $\Theta_{\alpha}$:
\eq{
  \arraycolsep2pt
 \begin{array}{lcl}
 \displaystyle   \phi_\alpha&\rightarrow&\displaystyle  \phi_\alpha+\Theta_\alpha \,,
 \\[4pt]
 \displaystyle  \omega_\alpha&\rightarrow& \displaystyle   \omega_\alpha-\Theta_\alpha \,.
  \end{array}
}
  
\end{enumerate}
Note that for Dirichlet boundary conditions the boundary term $\Omega_{\partial\Sigma}$ vanishes,
and therefore the last two symmetries are slightly modified and less restrictive.


\subsection{Recovering the ungauged world-sheet action}
\label{sec_recovery}

We finally want to show how the ungauged world-sheet theory \eqref{0}
can be recovered from the  gauged action \eqref{10}. 
This will be done using the equations of motion of the Lagrange multipliers $\chi_{\alpha}$
(and $\phi_{\alpha}$) \cite{Rocek:1991ps,Giveon:1993ai,Alvarez:1993qi}.


\subsubsection*{Equations of motion for $\chi_{\alpha}$ (and $\phi_{\alpha}$)}

Let us start by determining the equations of motion for the Lagrange multipliers
from the gauged action \eqref{10}. We distinguish again between all-Dirichlet or all-Neumann boundary conditions:
\begin{itemize}

\item In the case of Dirichlet boundary conditions, we recall from \eqref{dirc_7347375} that 
the one-form $\Omega_{\partial\Sigma}$ vanishes. The variation of the action \eqref{10} 
with respect to  $\chi_{\alpha}$
then leads to 
\eq{
\label{31}
\delta_\chi \widehat{\mathcal S} = \frac{i}{2\pi\alpha'} \int_{\Sigma} 
\delta \chi_{\alpha} \left( \der A^\alpha-\frac{1}{2}\op f_{\beta\gamma}{}^\alpha A^\beta\wedge A^\gamma 
\right)
,
}
where the boundary term vanishes due to the Dirichlet conditions \eqref{bc_local_2}.
Setting to zero the variation \eqref{31} leads to the equations of motion for $A^{\alpha}$, and together with 
the boundary condition on $A^{\alpha}$ we have
\eq{
\label{37}
 0= F^{\alpha} = 
 \der A^\alpha-\frac{1}{2}\op f_{\beta\gamma}{}^\alpha A^\beta\wedge A^\gamma 
 \,,
 \hspace{50pt}
  0 = A^{\alpha}\, \Bigr\rvert_{\partial\Sigma} \,.
}

\item Next, we turn to the Neumann boundary conditions. In this case the boundary one-form 
$\Omega_{\partial\Sigma}$ takes the form given in \eqref{neum_2846}, and for the equations 
of motion for $\chi_{\alpha}$ and $\phi_{\alpha}$ we determine
\eq{
\arraycolsep2pt
\begin{array}{lcll}
\displaystyle \delta_\chi \widehat{\mathcal S} &=& \displaystyle \frac{i}{2\pi\alpha'} \int_{\Sigma} 
& \displaystyle 
\delta \chi_{\alpha} \left( \der A^\alpha-\frac{1}{2}\op f_{\beta\gamma}{}^\alpha A^\beta\wedge A^\gamma 
\right) ,
\\[12pt]
\displaystyle \delta_\phi \widehat{\mathcal S} &=& \displaystyle  \frac{i}{2\pi\alpha'} \int_{\partial\Sigma} 
& \displaystyle 
\delta \phi_{\alpha} \, A^{\alpha}\,,
\end{array}
}
which gives the equations of motion 
\eq{
\label{37b}
 0= F^{\alpha} = 
 \der A^\alpha-\frac{1}{2}\op f_{\beta\gamma}{}^\alpha A^\beta\wedge A^\gamma 
 \,,
 \hspace{50pt}
  0 = A^{\alpha}\, \Bigr\rvert_{\partial\Sigma} \,.
}
Here it also becomes apparent why in the case of Neumann boundary conditions we introduced 
a second set of Lagrange multipliers $\phi_{\alpha}$. The latter are needed in order to set 
to zero the gauge field on the boundary \cite{Forste:1996hy}.

\end{itemize}


\subsubsection*{Abelian isometry algebra}

Let us now consider abelian isometry algebras for which the structure constants
$f_{\alpha\beta}{}^{\gamma}$ are vanishing. In this case the equation of motion 
for $\chi_{\alpha}$ imply that $A^{\alpha}$ is closed, and according to \eqref{hodge_decomp} we can therefore decompose
\eq{
  \label{exp_aa}
  A^{\alpha} = \der a_{(0)}^{\alpha} + \sum_{\mathsf m} a^{\alpha}_{(\mathsf m)} \op \varphi^{\mathsf m} \,,
}
where $a^{\alpha}_{(0)}$ are globally defined functions on $\Sigma$, $a^{\alpha}_{(\mathsf m)}\in\mathbb R$ 
and $\varphi^{\mathsf m}\in CcC_N^1$ is a basis of closed and co-closed one-forms on $\Sigma$ whose normal 
part vanishes. 
However, taking into account also 
the second condition in \eqref{37} and \eqref{37b}
we see that the tangential part of $A^{\alpha}$ 
is required to vanish on the boundary. As we discussed before, this implies that the $CcC^1_N$ part of
$A^{\alpha}$ is trivial, i.e. $a_{(\mathsf m)}^{\alpha} = 0$. 
Therefore $A^{\alpha}$ is exact -- and using the gauge symmetry \eqref{15} we can 
set $A^{\alpha}$ to zero. We have then recovered the original action \eqref{0}
from the gauged one \eqref{10}.


\subsubsection*{Non-abelian isometry algebra}

In the case of a non-abelian isometry algebra the situation is different. Since due to 
the first condition in 
\eqref{37} and \eqref{37b} the
gauge fields $A^{\alpha}$ are not closed, we cannot apply the Hodge decomposition theorem.
Heuristically, we can follow a method similar to the one in \cite{Plauschinn:2014nha}
where a field redefinition from $\DX{i} = \dx{i} + k^i_{\alpha} A^{\alpha}$
to $\der Y^i$ has been discussed. This procedure allows to 
recover the original action from the gauged action also in the case of non-abelian isometries, 
however, it does not take into account a non-trivial topology of
the world-sheet.

More accurate would be to start from the cohomology of the gauge-covariant derivative
and determine a corresponding Hodge decomposition theorem for manifolds with 
boundary. This is however beyond the scope of this paper.


\section{T-duality}
\label{sec_t-duality}

In this section we discuss collective T-duality transformations for the open string. 
In section~\ref{sec_t_gen} we first present results for the closed-string sector of the T-dual theory, 
whereas in section~\ref{sec_t_neu} and \ref{sec_t_dir}
we consider the open-string sector with  Neumann 
and  Dirichlet boundary conditions.


\subsection{Closed-string sector}
\label{sec_t_gen}

We start by determining the metric and Kalb-Ramond
$B$-field of the T-dual background. We do so by following Buscher's procedure \cite{Buscher:1987sk,Buscher:1987qj} of 
gauging target-space isometries --  as discussed in section~\ref{sec_gauging_ws} -- and integrating-out the 
corresponding gauge fields $A^{\alpha}$.


\subsubsection*{Equations of motion for $A^\alpha$}

The equations of motion for the gauge fields $A^{\alpha}$ are obtained by
varying the gauged action \eqref{10} with respect to $A^{\alpha}$. Since
the latter appear without a derivative, we can solve the equations of motion
algebraically. From the part of the action defined on the bulk $\Sigma$ we find using matrix notation
\cite{Plauschinn:2014nha}
\eq{
 \label{eomsigma}
 A^\alpha=-\Bigl( \op \bigl[\op \mathcal{G}-{\mathcal{D}}\op\mathcal{G}^{-1}
 {\mathcal{D}}\bigr]^{-1}\Bigr)^{\alpha\beta} \bigl( \op \mathds 1+i\star {\mathcal{D}} \op
 \mathcal{G}^{-1}\bigr)_\beta^{\hspace{6pt}\gamma} \, \bigl( \op\mathsf k+i\star {\xi} \op \bigr)_\gamma\,,
}
where we recall that $\alpha,\beta=1,\ldots,N$. For ease of notation,  we have defined the 
following quantities
\eq{
  \label{ioq_001}
  \arraycolsep2pt
 \begin{array}{lcl@{\hspace{60pt}}lcl}
 \displaystyle \mathcal{G}_{\alpha\beta} &=& \displaystyle k^i_\alpha \op G_{ij} \op k^j_\beta\,, 
 &
 \displaystyle {\xi}_\alpha &=& \displaystyle \der \chi_\alpha+\tilde{v}_\alpha\,, 
 \\[8pt]
 \displaystyle {\mathcal{D}}_{\alpha\beta} &=& \displaystyle \iota_{k_{[\underline{\alpha}}}
 \tilde{v}_{\underline{\beta}]}+f_{\alpha\beta}{}^\gamma \chi_\gamma\,, 
 &
 \displaystyle  \mathsf k_\alpha &=& \displaystyle k^i_\alpha \op G_{ij}\op\dx{j} \,.
 \end{array}
}
Note that although in \eqref{eomsigma} the inverse of the matrix $\mathcal G$ appears, 
in the integrated-out action only the inverse of $(\mathcal G \pm \mathcal D)$ plays a role. 
We therefore require invertibility only for the latter.

The contribution to the equations of motion for $A^{\alpha}$ from the boundary $\partial\Sigma$ 
depends on the type of boundary conditions for the gauge fields \eqref{bc_local_2}. For instance, 
if we impose Dirichlet conditions the 
gauge fields are absent on the boundary as shown in \eqref{dirc_7347375}. 
On the other hand, if $A^{\alpha}$ satisfy 
Neumann boundary conditions we find a non-trivial condition which has to be imposed as a constraint.
In particular, we have
\eq{
\label{eompsigma}
  \arraycolsep2pt
  \begin{array}{@{}l@{\hspace{40pt}}l@{}}
  \mbox{Dirichlet} & \displaystyle \varnothing \,,
  \\[6pt]
  \mbox{Neumann} & \displaystyle 0 = 2\pi\alpha' \op \iota_{k_{\alpha}} a-(\chi_\alpha+ \phi_\alpha + \omega_\alpha) \,\Bigr\rvert_{\partial\Sigma}\,.
  \end{array}
}


\subsubsection*{Integrated-out action}

Using the expressions \eqref{eomsigma} and
\eqref{eompsigma}, we can now evaluate the action \eqref{10}.
We obtain the following general form
\eq{
\label{ioaction}
\arraycolsep2pt
 \begin{array}{lcll}
 \displaystyle \check{\mathcal S}&=&\displaystyle -\frac{1}{2\pi\alpha'}\int_{\Sigma}&\displaystyle 
  \biggl[ \op \check{G}+
 i\check{B}+\frac{\alpha'}{2}\op\mathsf R\, \phi\star 1 \biggr]  
 \\[12pt]
  &&\displaystyle -\frac{1}{2\pi\alpha'}\int_{\partial\Sigma}  &
 \displaystyle \biggl[ \,2\pi i\op \alpha'\op a_a\op
 \dx{a}+ \alpha' \op k(s) \op
   \phi \, \der s \biggr]\,,
   \end{array}
}
with the world-sheet quantities $\check{G}$ and $\check{B}$ 
given by the  expressions
\eq{
\label{etargspgb2}
\arraycolsep2pt
\begin{array}{lclrr}
 \check{G}&=&\displaystyle G-  \frac{1}{2}
   &\displaystyle (\mathsf k+{\xi})^{T}\, \bigl(\mathcal{G}+{\mathcal{D}}\bigr)^{-1}\,\wedge
   &\displaystyle \star (\mathsf k-{\xi})\,,
 \\[14pt]
 \check{B}&=&\displaystyle B- \frac{1}{2}
   &\displaystyle(\mathsf k+{\xi})^{T}\,\bigl(\mathcal{G}+{\mathcal{D}}\bigr)^{-1}\,\wedge
   & (\mathsf k-{\xi})\,,
\end{array}
}
in which matrix multiplications for the indices $\alpha,\beta=1,\ldots,N$ is understood. 
We note that the original metric  and $B$-field appearing in \eqref{etargspgb2} read 
$G = \frac12 G_{ij} \op \dx{i}\wedge\star\dx{j}$
and $B = \frac12\op B_{ij} \op\dx{i}\wedge\dx{j}$. 
The fields in \eqref{etargspgb2} can be regarded as a  ``metric and $B$-field''
for an {\em enlarged target space} of dimension $D+N$, which is locally parametrized by 
coordinates $\{X^i, \chi_\alpha\}$ \cite{Plauschinn:2013wta}.


\subsubsection*{Change of basis}

The symmetric matrix $\check G$ defined 
through \eqref{etargspgb2} has $N$ null-eigenvalues. 
In the basis $\{\dx{i},\der \chi_\alpha\}$ the corresponding $N$ null-eigenvectors are
of the form \cite{Plauschinn:2013wta,Plauschinn:2014nha}
\eq{
\label{eigenvector}
\check{n}_\alpha=\begin{pmatrix}
                 k^i_\alpha \\
                 \mathcal{D}_{\alpha\beta}-\iota_{k_{\alpha}} \tilde v_{\beta}
                 \end{pmatrix} \,,
}
which can be used to perform a change of basis. 
Since the Killing vectors $k_{\alpha}$ are assumed to be linearly independent, we 
can always find a coordinate system in which the 
$N\times N$ matrix $ k_{\alpha}^{\beta}$ is invertible and where all other components of $k^i_{\alpha}$ 
vanish.\footnote{This is true when the isometry group has no fixed points or 
only isolated fixed points. If however the isotropy of the isometry group is non-trivial, 
the matrix $ k_{\alpha}^{\beta}$ is not invertible \cite{Giveon:1993ai} and a different basis of one-forms 
$e^{\alpha}$, $e^m$ and $e_{\alpha}$ in \eqref{basis_11}
has 
to be chosen. \label{foot_1}
}
Let us then define the following basis of one-forms
\eq{
  \label{basis_11}
  \arraycolsep2pt
  \begin{array}{@{\hspace{-10pt}}lclcl@{\hspace{-20pt}}}
  \displaystyle e^{\alpha} &=& \displaystyle  \bigl( k^{-1}\bigr)^{\alpha}_{\beta} \,\dx{\beta} \,,
  \\[12pt]
  \displaystyle e^m &=& \displaystyle \dx{m}  \,,
  \\[8pt]
  \displaystyle e_{\alpha} &=& \displaystyle  \der\chi_{\alpha} + \bigl[ \iota_{k_{(\ov\alpha}} v_{\ov\beta)} + f_{\alpha\beta}{}^{\gamma}  \chi_{\gamma} \bigr] \op \bigl( k^{-1}\bigr)^{\beta}{}_{\gamma} \,\dx{\gamma} 
  \,,
  \end{array}
}
where the indices take values 
$\alpha,\beta= 1,\ldots, N$ and $m,n=N+1,\ldots, D$.
We can now express the fields \eqref{etargspgb2} in this new basis:
\begin{itemize}

\item Since the symmetric two-tensor $\check G$ has $N$ zero-eigenvalues, it can be brought 
into the form
\eq{
    \check G = \frac{1}{2} \, \check{\mathsf G}_{IJ}\op e^I\wedge \star e^J \,,
}
where we employed the notation $e^I = \{e_{\alpha},e^m\}$ with $I=1,\ldots,D$.
Using the definitions shown in \eqref{ioq_001} and \eqref{def_v_002},
the components $\check{\mathsf G}_{IJ}$ take the following form
\eq{
  \label{dual_g_007}
  &  \arraycolsep2pt\check{\mathsf G}_{mn} = \begin{array}[t]{lcll@{\hspace{2pt}}c@{\hspace{2pt}}ll}
  G_{mn} 
  & - & \mathsf k_{\alpha m} 
  & \displaystyle \bigl[ (\mathcal G + \mathcal D)^{-1}& \displaystyle \mathcal G &\displaystyle (\mathcal G - 
  \mathcal D)^{-1} \bigr]^{\alpha\beta} & \mathsf k_{\beta n}
  \\[2pt]
  & - & \mathsf k_{\alpha m} 
  & \displaystyle \bigl[ (\mathcal G + \mathcal D)^{-1}& \displaystyle \mathcal D &\displaystyle (\mathcal G - 
  \mathcal D)^{-1} \bigr]^{\alpha\beta} & \tilde v_{\beta n}
  \\[2pt]
  & + & \tilde v_{\alpha m} 
  & \displaystyle \bigl[ (\mathcal G + \mathcal D)^{-1}& \displaystyle \mathcal D &\displaystyle (\mathcal G - 
  \mathcal D)^{-1} \bigr]^{\alpha\beta} & \mathsf k_{\beta n}
  \\[2pt]  
  & + & \tilde v_{\alpha m} 
  & \displaystyle \bigl[ (\mathcal G + \mathcal D)^{-1}& \displaystyle \mathcal G &\displaystyle (\mathcal G - 
  \mathcal D)^{-1} \bigr]^{\alpha\beta} & \tilde v_{\beta n} 
  \end{array}
  \\[10pt]
  &  \arraycolsep2pt\check{\mathsf G}^{\alpha}{}_n = \begin{array}[t]{cl@{\hspace{2pt}}c@{\hspace{2pt}}ll}
  +& \displaystyle \bigl[ (\mathcal G + \mathcal D)^{-1}& \displaystyle \mathcal D &\displaystyle (\mathcal G - 
  \mathcal D)^{-1} \bigr]^{\alpha\beta} & \mathsf k_{\beta n}
  \\[2pt]
  +& \displaystyle \bigl[ (\mathcal G + \mathcal D)^{-1}& \displaystyle \mathcal G &\displaystyle (\mathcal G - 
  \mathcal D)^{-1} \bigr]^{\alpha\beta} & \tilde v_{\beta n}
  \end{array}
  \\[10pt]
  &\arraycolsep2pt \check{\mathsf G}^{\alpha\beta} =  \begin{array}[t]{cl@{\hspace{2pt}}c@{\hspace{2pt}}l}
  +&\displaystyle \bigl[ (\mathcal G + \mathcal D)^{-1}& \displaystyle \mathcal G &\displaystyle (\mathcal G - 
  \mathcal D)^{-1} \bigr]^{\alpha\beta}
  \end{array}
}
These expressions are the components of the metric of the dual background after performing 
a collective T-duality transformation along $N$ directions. 
In particular, for the case of a T-duality along one direction these formulas reduce to the 
usual Buscher rules \cite{Buscher:1987sk}.

\item For the two-form $\check B$ given in \eqref{etargspgb2}
we find a slightly different structure. In particular, we can write
\eq{
  \label{dual_004}
  \check B = \frac{1}{2} \, \check{\mathsf B}_{IJ}\op e^I\wedge e^J + \check B^{\rm res.} \,,
}
where the anti-symmetric matrix $\check{\mathsf B}_{IJ}$ takes the form
\eq{
  \label{dual_b_001}
  &  \arraycolsep2pt\check{\mathsf B}_{mn} = \begin{array}[t]{lcll@{\hspace{2pt}}c@{\hspace{2pt}}ll}
  B_{mn} 
  & + & \mathsf k_{\alpha m} 
  & \displaystyle \bigl[ (\mathcal G + \mathcal D)^{-1}& \displaystyle \mathcal D &\displaystyle (\mathcal G - 
  \mathcal D)^{-1} \bigr]^{\alpha\beta} & \mathsf k_{\beta n}
  \\[2pt]
  & + & \mathsf k_{\alpha m} 
  & \displaystyle \bigl[ (\mathcal G + \mathcal D)^{-1}& \displaystyle \mathcal G &\displaystyle (\mathcal G - 
  \mathcal D)^{-1} \bigr]^{\alpha\beta} & \tilde v_{\beta n}
  \\[2pt]
  & - & \tilde v_{\alpha m} 
  & \displaystyle \bigl[ (\mathcal G + \mathcal D)^{-1}& \displaystyle \mathcal G &\displaystyle (\mathcal G - 
  \mathcal D)^{-1} \bigr]^{\alpha\beta} & \mathsf k_{\beta n}
  \\[2pt]  
  & - & \tilde v_{\alpha m} 
  & \displaystyle \bigl[ (\mathcal G + \mathcal D)^{-1}& \displaystyle \mathcal D &\displaystyle (\mathcal G - 
  \mathcal D)^{-1} \bigr]^{\alpha\beta} & \tilde v_{\beta n} 
  \end{array}
  \\[10pt]
  &  \arraycolsep2pt\check{\mathsf B}^{\alpha}{}_n = \begin{array}[t]{cl@{\hspace{2pt}}c@{\hspace{2pt}}ll}
  -& \displaystyle \bigl[ (\mathcal G + \mathcal D)^{-1}& \displaystyle \mathcal G &\displaystyle (\mathcal G - 
  \mathcal D)^{-1} \bigr]^{\alpha\beta} & \mathsf k_{\beta n}
  \\[2pt]
  -& \displaystyle \bigl[ (\mathcal G + \mathcal D)^{-1}& \displaystyle \mathcal D &\displaystyle (\mathcal G - 
  \mathcal D)^{-1} \bigr]^{\alpha\beta} & \tilde v_{\beta n}
  \end{array}
  \\[10pt]
  &\arraycolsep2pt \check{\mathsf B}^{\alpha\beta} =  \begin{array}[t]{cl@{\hspace{2pt}}c@{\hspace{2pt}}l}
  -&\displaystyle \bigl[ (\mathcal G + \mathcal D)^{-1}& \displaystyle \mathcal D &\displaystyle (\mathcal G - 
  \mathcal D)^{-1} \bigr]^{\alpha\beta}
  \end{array}
}
These expressions give the $B$-field of the T-dual background, which in the case of a single T-duality
again match with the Buscher rules.

\end{itemize}
Let us finally address the residual $B$-field $\check B^{\rm res.}$ mentioned in \eqref{dual_004}. 
Through the one-forms $e^{\alpha}$ it depends on  $\der X^{\alpha}$ 
of the original background,  
and it takes the explicit form
\eq{
  \label{dual_bres}
  \check B^{\rm res.} =  \displaystyle e^{\alpha} \wedge \left[ \,  \der\chi_{\alpha} + v_{\alpha} +\frac12 
  \bigl( \iota_{k_{[\underline{\alpha}}} v_{\underline{\beta}]} + f_{\alpha\beta}{}^{\gamma}
  \chi_{\gamma} \bigr)\, e^{\beta}\,\right].
}
We discuss this expression separately for Neumann and Dirichlet boundary conditions 
in the following two subsections.


\subsubsection*{Dilaton}

The dual dilaton $\check \phi$ has to be determined by a one-loop computation as 
in \cite{Buscher:1987qj}. However, here we determine $\check \phi$ by demanding that
the combination $e^{-2\phi}\sqrt{\det G}$ is invariant under T-duality transformations. 
We then find
\eq{
  \label{dilaton}
  \check \phi = \phi - \frac{1}{4}\op \frac{\det G}{\det \check{\mathsf G}} \,.
}


\subsubsection*{Closure of basis}

In our above discussion, we have identified $e^I = \{e_{\alpha},e^m\}$ with $I=1,\ldots,D$
as the basis one-forms of the T-dual background. As such, they have to be closed under the 
exterior derivative $\der$. 
Let us therefore compute the following expressions
\eq{
  \label{dual_forms_824}
  \arraycolsep2pt
  \begin{array}{@{\hspace{-10pt}}lcl@{\hspace{-30pt}}}
  \displaystyle \der e^{\alpha} &=& -\displaystyle  \frac{1}{2}\, f_{\beta\gamma}{}^{\alpha} e^{\beta}
  \wedge e^{\gamma} 
  - \bigl( k^{-1} \bigr)^{\alpha}_{\beta}\, \bigl[ \partial_m \op k^{\beta}_{\gamma} \bigr]\op e^m \wedge e^{\gamma}
  \,,
  \\[12pt]
  \displaystyle \der e^m &=& \,0\,,
  \\[8pt]
  \displaystyle  \der e_{\alpha} &=& \displaystyle - f_{\alpha\beta}{}^{\gamma} \op  e^{\beta} \wedge e_{\gamma}
  + \Bigl( \partial_m \op\iota_{k_{(\ov\alpha}} v_{\ov\beta)} 
 - \bigl[ \iota_{k_{(\ov\alpha}} v_{\ov\gamma)} + f_{\alpha\gamma}{}^{\delta}  \chi_{\delta} \bigr] 
  \bigl( k^{-1} \bigr)^{\gamma}_{\epsilon}\op \bigl[ \partial_m \op k^{\epsilon}_{\beta} \bigr]\Bigr)\op e^m \wedge e^{\beta}
  \,,
  \end{array}
}
from which we see that in general the basis $e^I = \{e_{\alpha},e^m\}$ does not close under the exterior derivative $\der$.  
This means that the dual background may implicitly depend on the original coordinates, 
which is a property expected from a non-geometric background. 
Nevertheless, if we restrict ourselves to either of the following situations
\eq{
  \label{cond_closure}
  \left\{ 
  \arraycolsep2pt
  \renewcommand{\arraystretch}{1.2}
  \begin{array}{lcl}
  0&=& \displaystyle f_{\alpha\beta}{}^{\gamma} 
  \\
  0&=& \displaystyle  \partial_m \op\iota_{k_{(\ov\alpha}} v_{\ov\beta)} 
  \\
  0&=& \displaystyle  \partial_m \op k_{\alpha}^{\beta}
  \end{array}
  \right\}\,,
  \hspace{50pt}
  \left\{ 
  \arraycolsep2pt
  \renewcommand{\arraystretch}{1.2}
  \begin{array}{lcl}
  0&=& \displaystyle f_{\alpha\beta}{}^{\gamma} 
  \\
  0&=& \displaystyle  \iota_{k_{(\ov\alpha}} v_{\ov\beta)} 
  \end{array}
  \right\}\,,  
}
we see that the basis $e^I = \{e_{\alpha},e^m\}$ is closed under the exterior derivative. 
Note also that, as mentioned in footnote~\ref{foot_1}, in the case of a non-trivial isotropy of the 
isometry group a different basis of one-forms has to be chosen. In this case the 
exterior algebra may take a different form.


\subsubsection*{Remark on non-geometric fluxes}

Given the general form of the dual metric and $B$-field after a collective T-duality 
transformation shown in \eqref{dual_g_007} and \eqref{dual_b_001}, we 
want to take the opportunity and remark on possible non-geometric fluxes. 
In particular, given $\check{\mathsf G}_{IJ}$ and $\check{\mathsf B}_{IJ}$ we can 
define a new metric $\mathsf g^{IJ}$ and  bi-vector field $\beta^{IJ}$ 
via
\eq{
  \bigl( \check{\mathsf G} \pm \check{\mathsf B} \bigr)^{-1} = 
  \mathsf g \pm \beta \,,
}
where $\mathsf g^{IJ}$ corresponds to the symmetric part and $\beta^{IJ}$ to the 
anti-symmetric part. 
The non-geometric $Q$- and $R$-fluxes are then expressed in 
terms of $\beta$ as follows
\eq{
  \label{def_qr}
  Q_I{}^{JK} = \partial_I \beta^{JK} \,,
  \hspace{60pt}
  R^{IJK} = 3\op \beta^{[\ul I M} \partial_M \beta^{\ul J \ul K]} 
  \,.
}

Let us now consider a $D$-dimensional background and perform a collective
T-duality transformation along all $D$ directions. Ignoring for a moment
that the basis of dual one-forms does not close among itself under $\der$,
from \eqref{dual_g_007} and \eqref{dual_b_001} we can determine 
\eq{
  \mathsf g_{\alpha\beta} = \mathcal G_{\alpha\beta} \,,
  \hspace{60pt}
  \beta_{\alpha\beta} = \mathcal D_{\alpha\beta} \,,
}
where we note that for the dual coordinates $\chi_{\alpha}$ the position of 
the index is reversed as compared to the original coordinates $X^{\alpha}$.
For the non-geometric fluxes we  compute using \eqref{16_1}
and the Jacobi identity of $f_{\alpha\beta}{}^{\gamma}$
\eq{
  \label{dual_fluxes_24}
  Q^{\alpha}{}_{\beta\gamma} = f_{\beta\gamma}{}^{\alpha} \,,
  \hspace{60pt}
  R_{\alpha\beta\gamma} = \iota_{k_{\alpha}}\iota_{k_{\beta}}\iota_{k_{\gamma}} H \,.
}
The general form of these expressions  is as expected: for a background with a non-abelian isometry 
group the metric is usually non-trivial and one expects a corresponding non-geometric flux
related to $f_{\alpha\beta}{}^{\gamma}$. Under a collective T-duality transformation 
along all directions this geometric flux should be mapped into the non-geometric $Q$-flux, 
as in \eqref{dual_fluxes_24}. 
Furthermore, under a collective T-duality transformation along all directions the 
$H$-flux is expected to be mapped into the non-geometric $R$-flux, as can be seen 
in  \eqref{dual_fluxes_24}.
However, as mentioned above, the dual-basis one-forms $e_{\alpha}$ defined in
\eqref{basis_11} do not closed under the exterior derivative 
\eq{
  \der e_{\alpha} = - f_{\alpha\beta}{}^{\gamma} \op  e^{\beta} \wedge e_{\gamma} \,.
}
In this way the dual geometry implicitly depends on the original coordinates, 
which on general grounds is expected from a non-geometric background.


\subsection{Open-string sector -- Neumann directions}
\label{sec_t_neu}

Let us now consider the open-string sector and specialize to T-duality transformations
along multiple Neumann directions $X^a$.
T-duality along Dirichlet directions will be discussed  in section~\ref{sec_t_dir}, but the 
mixed case of collective T-duality along Neumann and Dirichlet directions at the same time will not be studied
separately.


\subsubsection*{Integrating-out I -- gauge fields $A^{\alpha}$}

We start by integrating-out the gauge fields $A^{\alpha}$ from the gauged action \eqref{10}, 
taking into account the  Neumann boundary conditions shown in \eqref{bc_local_2}.
The solution to the equations of motion for $A^{\alpha}$ in $\Sigma$ has been given in equation \eqref{eomsigma},
which leads to the dual metric and $B$-field shown in \eqref{dual_g_007} and 
\eqref{dual_b_001}, including the residual $B$-field \eqref{dual_bres}.
The contribution to the equation of motion for $A^{\alpha}$ coming from the boundary leads to the constraint \eqref{eompsigma}, which we
implement as a $\delta$-function into the path integral
\eq{
  \label{dual_delta_111}
  \delta\bigl( \phi_{\alpha} - \tilde\chi_{\alpha}  \bigr)_{\partial\Sigma} \,,
   \hspace{50pt} \tilde \chi_{\alpha} = \chi_{\alpha} + \omega_{\alpha} - 2\pi\alpha' \iota_{k_{\alpha}} a\,.
}
Furthermore, the gauge fields $A^{\alpha}$ are subject to the Neumann boundary conditions. 
Evaluating these for the solution \eqref{eomsigma} gives the 
following general condition on the boundary
\eq{
\label{neum_bc_a_23}
0 = \left[ 
  G_{ai}\op k^i_{\alpha} \op\star A^{\alpha}\bigr\rvert_{\eqref{eomsigma}}+ 2\pi \alpha' i \op \mathcal F_{ab} \op 
  k_{\alpha}^b
  \op A^{\alpha}\bigr\rvert_{\eqref{eomsigma}} \right]_{\partial\Sigma}\,,
}
where $A^{\beta}\rvert_{\eqref{eomsigma}}$ denotes the solution \eqref{eomsigma}.
In general \eqref{neum_bc_a_23} will take a complicated form and has to 
be computed in a case-by-case analysis. 
However, if we restrict ourselves for a moment to the abelian situation with $f_{\alpha\beta}{}^{\gamma}=0$ 
and contract \eqref{neum_bc_a_23} with $k_{\beta}$, we find that 
\eq{
  0 =  \der \tilde\chi_{\alpha} \bigr\rvert_{\partial\Sigma}
 \,.
}
These relations describe Dirichlet boundary conditions for the dual coordinates $\chi_{\alpha}$, which 
we expect on general grounds.


\subsubsection*{Integrating-out II -- Lagrange multipliers $\phi_{\alpha}$}

Next, we consider the Lagrange multipliers $\phi_{\alpha}$. After integrating-out the gauge fields $A^{\alpha}$
and implementing the constraint \eqref{dual_delta_111}, 
the path integral takes the following schematic form
\eq{
  \label{path_90284}
  \mathcal Z = \int \frac{[\mder X^i]\op [\mder \chi_{\alpha}]}{\mathcal V_{\rm gauge}}
  \int [\mder \phi_{\alpha}]\: \delta\bigl( \phi_{\alpha} - \tilde\chi_{\alpha}  \bigr)_{\partial\Sigma} \:
  \exp\Bigl( \check{\mathcal S}[X^i,\chi_{\alpha}]\Bigr) \,,
}
where $\mathcal V_{\rm gauge}$ denotes the volume of the local gauge symmetry \eqref{15},
$\tilde \chi_{\alpha}$ have been defined in \eqref{dual_delta_111}
and $\check{\mathcal S}$ denotes the action \eqref{ioaction}.
Since the latter does not depend on $\phi_{\alpha}$, the integral over $\phi_{\alpha}$ can 
performed trivially and the $\delta$-function \eqref{dual_delta_111} gives one.


\subsubsection*{Integrating-out III -- coordinates $X^{\alpha}$}

The action $\check{\mathcal S}$ still depends on the original coordinates $X^{\alpha}$ which satisfy Neumann boundary
conditions. This means in particular that $\der X^{\alpha}$ can have a non-vanishing 
$CcC_N^1$-part, so the local symmetry \eqref{15} cannot be used to set $X^{\alpha}$ to zero. 
However, the residual $B$-field \eqref{dual_bres} provides the required terms.
In the following we restrict ourselves again to the abelian situation and 
make the technical assumption that $k_{\alpha}^{\beta}$ are constant, but more general 
cases can be treated in a similar fashion.

To start, let us note that in \eqref{constr1}  we have shown conditions which relate 
the open-string gauge field $a$, the one-forms $ v_{\alpha}$ and the functions $\omega_{\alpha}$ 
on the boundary $\partial\Sigma$ to each other. All these quantities depend only on $X^i$,
which have a unique continuation from the boundary $\partial\Sigma$ to the bulk $\Sigma$. 
We can therefore assume  that the relations \eqref{constr1} and \eqref{16_1}
are valid also on $\Sigma$. 
This allows us to rewrite $\check B{}^{\rm res.}$ in the following way
\eq{
  \check B^{\rm res.} = \der \Bigl[ - \tilde\chi_{\alpha} \op
  e^{\alpha} 
  - 2\pi \alpha' a \Bigr] + 2\pi\alpha' \bigl(\op\tfrac{1}{2}\op F_{mn}e^m\wedge e^n \bigr)\,,
}
with $F=\der a$ the 
open-string field strength and $\tilde\chi_{\alpha}$ were defined in \eqref{dual_delta_111}.
The dual action \eqref{ioaction} contains $\check B{}^{\rm res.}$ together with the 
open-string gauge field  on the boundary. For those we compute
\begin{align}
\nonumber
 &-\frac{i}{2\pi\alpha'}\int_{\Sigma} \check B^{\rm res.} 
 -\frac{i}{2\pi\alpha'}\int_{\partial\Sigma} 2\pi\alpha' a
 \\
  \label{exp_dual_2947}
 =&+\frac{i}{2\pi\alpha'}\int_{\partial\Sigma} \tilde\chi_{\alpha} \op 
  e^{\alpha} 
  -\frac{i}{2\pi\alpha'}\int_{\Sigma} 2\pi\alpha' \bigl(\tfrac{1}{2} F_{mn}\op e^m\wedge e^n \bigr) \,.
\end{align}
The second term in \eqref{exp_dual_2947} denotes the open-string field strength along the 
directions which are not dualized and combines with $B_{mn}$ in \eqref{dual_b_001} 
into the gauge-invariant open-string field strength. 
Turning to the first term in  \eqref{exp_dual_2947}, since the $\der X^{\alpha}$ with Neumann boundary conditions appearing in $e^{\alpha}$ are closed, we can expand them into an exact and a $CcC_N^1$-part 
similarly as in equation  \eqref{exp_aa}
\eq{
  \label{exp_xx}
  \der X^{\alpha} = \der X_{(0)}^{\alpha} + \sum_{\mathsf m} X^{\alpha}_{(\mathsf m)} \op \varphi^{\mathsf m} \,,
}
where \raisebox{0pt}[0pt][0pt]{$X_{(0)}^{\alpha}$} are globally-defined functions on $\Sigma$, 
\raisebox{0pt}[0pt][0pt]{$X^{\alpha}_{(\mathsf m)} \in\mathbb R$} 
are constants and $\varphi^{\mathsf m}\in CcC_N^1$ is a basis of closed and co-closed one-forms with 
vanishing normal part. 
A corresponding basis of the first homology
on $\partial\Sigma$ will be denoted by $\gamma_{\mathsf m}$ and can be normalized as 
\raisebox{0pt}[0pt][0pt]{$\int_{\gamma_{\mathsf m}} \varphi^{\mathsf n} = \delta_{\mathsf m}^{\mathsf n}$}.
Now, the exact part  \raisebox{0pt}[0pt][0pt]{$X_{(0)}^{\alpha}$} appearing in \eqref{exp_xx} 
can be set to zero using the symmetry \eqref{15},
while for the $CcC_N^1$-part we distinguish the following two situations:
\begin{itemize}

\item Let us first assume that we can define winding/momentum numbers 
$n_{(\mathsf m)}$ for the $X^{\alpha}$ as follows
\eq{
  \oint_{\gamma_{\mathsf m}} \der X^{\alpha} = X^{\alpha}_{(\mathsf m)} = 2\pi \op n^{\alpha}_{(\mathsf m)} \,,\hspace{50pt}
  n^{\alpha}_{(\mathsf m)} \in \mathbb Z\,,
}
which determine the \raisebox{0pt}[0pt][0pt]{$X^{\alpha}_{(\mathsf m)}$}  appearing in \eqref{exp_xx}. 
For a compactification of $X^{\alpha}$ on a circle or a flat torus without $H$-flux these 
momentum/winding sectors always exist, but on more general backgrounds 
these may be either absent or not be quantized. 
Coming now back to the first expression in \eqref{exp_dual_2947}, we see that the path integral 
\eqref{path_90284} (after integrating over $\phi_{\alpha}$) contains the following terms
\eq{
  \label{path_com_3987256}
    \mathcal Z &\supset \int \frac{[\mder X^{\alpha}]}{\mathcal V_{\rm gauge}} 
   \,\exp\left[
   \frac{i}{2\pi\alpha'}\int_{\partial\Sigma} \tilde\chi_{\alpha} \op   e^{\alpha} 
   \right]
   \\[2pt]
   &\supset \int \frac{\bigl[\mder X_{(0)}^{\alpha}\bigr]}{\mathcal V_{\rm gauge}} 
   \sum_{n^{\alpha}_{(\mathsf m)}\in\mathbb Z}
   \,\exp\left[
   \frac{i}{2\pi\alpha'}\int_{\partial\Sigma} \tilde\chi_{\beta} \op   \bigl(k^{-1}\bigr)_{\alpha}^{\beta} \,dX^{\alpha}
   \right]
   \\
   &\supset \sum_{n_{(\mathsf m)}^{\alpha} \in\mathbb Z}
   \exp\left[
   \frac{i}{\alpha'} \,\tilde\chi_{\beta} \op   \bigl(k^{-1}\bigr)_{\alpha}^{\beta}\,
   n^{\alpha}_{(\mathsf m)}
   \right]_{\partial\Sigma}
   \\
   &\supset \sum_{m_{\alpha(\mathsf m)} \in\mathbb Z} \delta
   \left[
   \frac{1}{2\pi \alpha'} \,\tilde\chi_{\beta} \op   \bigl(k^{-1}\bigr)_{\alpha}^{\beta}
   -
   m_{\alpha(\mathsf m)}
   \right]_{\partial\Sigma},
}
where from the second to the third line we set to zero the exact part using the local symmetries,  and from the third to the 
fourth line we employed the definition of the  periodic Kronecker $\delta$-symbol \cite{Rocek:1991ps}.
We therefore see that coordinates $\tilde \chi_{\alpha}$ on the boundary $\partial\Sigma$ are 
quantized as
\eq{
  \label{quant_dual_2}
  \frac{1}{\alpha'} \op \,\tilde\chi_{\beta} \op   \bigl(k^{-1}\bigr)_{\alpha}^{\beta} \op\Bigr\rvert_{\partial\Sigma}
  \in2\pi \mathbb Z\,.
}

\item The second possibility is that \raisebox{0pt}[0pt][0pt]{$X^{\alpha}_{(\mathsf m)}$} appearing in 
\eqref{exp_xx} are real numbers determined via the equations of motion \eqref{eom_x_001}, 
which are in particular not quantized. 
In this case the sum in \eqref{path_com_3987256} is replaced by integrals 
over \raisebox{0pt}[0pt][0pt]{$X^{\alpha}_{(\mathsf m)}$}, leading to Dirac $\delta$-functions 
which set
\eq{
  \label{quant_dual_3}
  \tilde\chi_{\alpha} \op\bigr\rvert_{\partial\Sigma}
  =0\,.
}

\end{itemize}
Let us also note that $\tilde\chi_{\alpha}$ defined in \eqref{dual_delta_111} contain a contribution 
from the open-string gauge field $a$. This means that a non-vanishing $\iota_{k_{\alpha}} a$ 
leads to a shift of the dual coordinates $\chi_{\alpha}$, which is again expected on general grounds.


\subsubsection*{Summary}

Let us summarize the main steps to obtain the dual open-string sector in the case of 
Neumann boundary conditions. We illustrated this procedure with 
an abelian isometry algebra with constant Killing vectors, but more general 
configurations (subject to the questions discussed above) follow a similar 
pattern:
\begin{enumerate}

\item First, we integrate-out the gauge fields $A^{\alpha}$ from the gauged action 
\eqref{10}.
This gives the dual metric and $B$-field in the bulk and imposes the constraint 
\eqref{dual_delta_111} on the boundary. 
The boundary conditions for $A^{\alpha}$ shown in \eqref{neum_bc_a_23} imply that the 
dual coordinates satisfy Dirichlet boundary conditions. 

\item Next, we integrate over the Lagrange multipliers $\phi_{\alpha}$ in the path integral. 
Due to the $\delta$-function \eqref{dual_delta_111} this integral gives one. 

\item Finally, we perform the path integral over the original coordinates $X^{\alpha}$. 
The exact part of $\der X^{\alpha}$ can be gauged to zero using the local symmetry
\eqref{15}, while the co-homologically non-trivial part of $\der X^{\alpha}$ 
appears in the residual $B$-field \eqref{dual_bres}.
The latter either gives rise to a periodic Kronecker $\delta$-symbol in the path integral 
leading to quantization conditions for the dual coordinates on the boundary,
or gives a Dirac $\delta$-function which imposes Dirichlet conditions for the dual 
coordinates.

\end{enumerate}
We want to point-out that these results are in agreement with the well-known CFT analysis of T-duality for the
open string: T-duality 
along a Neumann direction results in a dual Dirichlet direction, and
a non-trivial Wilson line leads to a shift of the dual coordinates on the boundary.


\subsection{Open-string sector -- Dirichlet directions}
\label{sec_t_dir}

We now turn to collective  T-duality transformations along directions with Dirichlet boundary conditions. 
Due to the absence of the Lagrange multipliers $\phi_{\alpha}$, the procedure differs slightly 
from the Neumann case.


\subsubsection*{Integrating-out I -- gauge fields $A^{\alpha}$}

We start again by integrating-out the gauge fields $A^{\alpha}$
from the gauged action \eqref{10}.
The equations of motion for $A^{\alpha}$ in the bulk $\Sigma$ lead to the solution 
\eqref{eomsigma},
which in turn gives  the dual metric and $B$-field  shown in \eqref{dual_g_007} and 
\eqref{dual_b_001}. The residual $B$-field can be found in \eqref{dual_bres}
and
-- as already summarized in \eqref{eompsigma} --
there are no additional conditions arising from 
the variation of the action with respect to $A^{\alpha}$ on the boundary.

The boundary conditions for 
$A^{\alpha}$ shown in \eqref{bc_local_2} require the gauge fields to vanish on the boundary. 
For the solution \eqref{eomsigma} this implies in particular that
\eq{
  \label{dirc_bound_cond}
  0 = \left[ \:A^{\alpha}\bigr\rvert_{\eqref{eomsigma}} \right]_{\partial\Sigma}\,,
}
which using \eqref{eomsigma} and the basis
$e^I = \{e_{\alpha},e^m\}$ given in  \eqref{basis_11} can be expressed 
as
\eq{
  \label{dir_dual_bc}
  0 = \check{\mathsf G}^{\alpha}{}_I\bigl(e^I\bigr)_{\rm norm} 
  + i\op \check{\mathsf B}^{\alpha}{}_I\bigl(e^I\bigr)_{\rm tan}  \,.
}
By comparing \eqref{dir_dual_bc} with \eqref{boundary_cond} we conclude that these relations describe 
Neumann boundary conditions for the dual coordinates,
which is again expected on general grounds.


\subsubsection*{Integrating-out II -- coordinates $X^{\alpha}$}

Next, we turn to the original coordinates $X^{\alpha}$ which appear in the action 
via the residual $B$-field. 
In the following we assume for simplicity that the Killing vectors are constant and that $v_{\alpha}=0$, 
but for more general configurations can be treated in a similar way.

Since the original coordinates $X^{\alpha}$ 
satisfy  Dirichlet boundary conditions, the one-forms $\der X^{\alpha}$ are exact on $\Sigma$.
This allows us to rewrite the residual $B$-field 
\eqref{dual_bres} in the following way
\eq{
  \check B^{\rm res.} = e^{\alpha} \wedge \der\chi_{\alpha} = \der \Bigl[ X^{\alpha} \,\bigl(k^{-1}\bigr)_{\alpha}^{\beta}
  \, e_{\beta} \Bigr] \,,
}  
which for the action implies
\eq{
  -\frac{i}{2\pi\alpha'}\int_{\Sigma} \check B^{\rm res.}  = 
  -\frac{i}{2\pi\alpha'}\int_{\partial\Sigma} 2\pi\alpha' \left[ 
  \frac{X^{\alpha}(k^{-1})_{\alpha}^{\beta}}{2\pi\alpha'}
  \, e_{\beta}\right].
}
We therefore see that the position of the D-brane in the original theory $X^{\alpha}\rvert_{\partial\Sigma}$
determines a constant gauge field 
for the T-dual theory
\eq{
  \label{dir_324545}
  \check{\mathsf a}^{\alpha}= \frac{1}{2\pi\alpha'} \bigl(k^{-1}\bigr)^{\alpha}_{\beta}\, X^{\beta}\op \Bigr\rvert_{\partial\Sigma}\,.
}  
Using the local gauge symmetry \eqref{15} we can then 
fix $X^{\alpha}$ in the bulk $\Sigma$ to a convenient value, and trivially perform the 
corresponding integration in the path integral. In this way we have then obtained the T-dual theory.


\subsubsection*{Summary}

Let us  briefly summarize the main steps for obtaining the dual background for a collective
T-duality transformation along Dirichlet directions.
\begin{enumerate}

\item We first integrate-out the gauge fields $A^{\alpha}$ form the gauged action \eqref{10}
and obtain the dual metric and $B$-field. The boundary conditions 
\eqref{dirc_bound_cond} for the gauge fields then lead to Neumann boundary conditions for the dual 
coordinates. 

\item In contrast to the case of T-duality transformations along Neumann directions, in the 
present situation there are no Lagrange multipliers $\phi_{\alpha}$ present. 

\item In the case of  an abelian isometry algebra and vanishing one-forms 
$v_{\alpha}$, we can rewrite the residual $B$-field \eqref{dual_bres}. The latter
then leads to a Wilson line along the dual directions determined by
the position of the original D-brane. 
Since the original one-forms $\der X^{\alpha}$ are exact, we can use the local gauge symmetry to 
fix them to a convenient value in the bulk $\Sigma$ (subject to the boundary conditions on $\partial\Sigma$).

\end{enumerate}
We want to point-out again that these results agree with the results expected from a CFT 
analysis on a background with constant metric and $B$-field. In particular, 
a T-duality along a Dirichlet direction leads to a Neumann boundary condition, 
and the position of the original D-brane corresponds to a constant gauge field in the dual theory.


\section{Examples -- three-torus with $H$-flux}
\label{sec_examples}

In this section we want to illustrate the formalism introduced above with the example 
of the three-torus with $H$-flux.
We discuss a number of different settings and
show explicitly that the results expected from toroidal compactifications with constant $B$-field 
are obtained also for non-trivial $B$-field.


\subsubsection*{Setup}

As a starting point we consider the background of a flat three-torus with $H$-flux and different types of D-branes.
We denote coordinates on the three-torus $\mathbb T^3$ by 
$X^i$ with $i=1,2,3$, and impose 
the identifications $X^i\sim X^i+2\pi$.
The corresponding basis for the co-tangent space is given by 
one-forms $\dx{i}$, and
the metric and $B$-field read
\eq{
  \label{metr_bf}
  \arraycolsep2pt
 {G}_{ij} = \begin{pmatrix}
                                                   R_1^2 & 0 & 0 \\
                                                   0 & R_2^2 & 0 \\
                                                   0 & 0 & R_3^2
                                                  \end{pmatrix}
\,, 
 \hspace{40pt} B = \frac{\alpha'}{2\pi} \, h\op X^3\,\dx{1}\wedge\dx{2}\,,
 \hspace{40pt} \phi = \phi_0 \,,
}
where $h\in\mathbb{Z}$ due to the flux-quantization condition. 
The radii $R_i$ have the dimension of the string-length $\ell_{\rm s}$, whereas
the coordinates $X^i$ are dimensionless. 
The dilaton $\phi$ is taken to be constant, and
the Killing vectors we are interested-in 
 (in a basis dual to $\dx{i}$) 
are given by 
\eq{
  \label{killingv}
  \arraycolsep2pt
 {k}_{1} = \begin{pmatrix}
                                                   1 \\
                                                   0 \\
                                                   0 
                                                  \end{pmatrix}
\,, 
\hspace{40pt}
{k}_{2} = \begin{pmatrix}
                                                   0 \\
                                                   1 \\
                                                   0 
                                                  \end{pmatrix}
\,,
\hspace{40pt}
{k}_{3} =\begin{pmatrix}
                                                   0 \\
                                                   0 \\
                                                   1 
                                                  \end{pmatrix}\,,
}
which satisfy an abelian isometry algebra.
Our conventions for the open-string sector is that a D$p$-brane has 
Neumann boundary conditions along the time direction and along $p$ spatial directions
in $\mathbb T^3$, while the remaining directions are of Dirichlet type.
Finally, since in our convention the coordinates $X^i$ are dimensionless
it turns out to be convenient to use also dimensionless dual coordinates
\eq{
\label{con_dual}
\check\chi_{\alpha} = \frac{1}{\alpha'} \,\chi_{\alpha} \,.
}


\subsection{One T-duality}
\label{sec:onetd}

We start with discussing one T-duality transformation for the above background. For convenience we 
always take the direction $X^1$, but the formalism introduced in section~\ref{sec_t-duality} 
gives similar results for the other directions. In particular, we 
can equally perform a T-duality transformation along the direction $X^3$.


\subsubsection*{D1-brane along $X^1$}

Let us place a D1-brane along the direction $X^1$ and consider a constant 
open-string gauge field $a$. The corresponding field strength $F=\der a$ vanishes, and 
the boundary conditions \eqref{boundary_cond}
take the form
\eq{
  0 = \bigl(\dx{1}\bigr)_{\rm norm} \,, \hspace{40pt}
  0 = \bigl(\dx{2}\bigr)_{\rm tan} \,, \hspace{40pt}
  0 = \bigl(\dx{3}\bigr)_{\rm tan} \,.
}
The constraints \eqref{constr1} and \eqref{16_1} 
for a T-duality along the $X^1$-direction are 
solved for instance by 
\eq{
  \arraycolsep2pt
  \begin{array}{lcl}
  a & = & \displaystyle  a_1 \op \dx{1} \,,
  \\[4pt]
  v_1 & =& \displaystyle  0\,,
  \\[4pt]
  \omega_1 &= &0 \,,
  \end{array}
  \hspace{80pt} a_1= {\rm const.}\,,
}
and the dual metric and $B$-field can be determined from 
the general expressions \eqref{dual_g_007} and \eqref{dual_b_001} (together with \eqref{con_dual}) as
\eq{
  \arraycolsep4pt
  \renewcommand{\arraystretch}{1.2}
  \check{\mathsf G}_{IJ} = 
  \left( \begin{array}{ccc}
        \frac{\alpha'^2}{R_1^2} & - \frac{\alpha'^2}{R_1^2} \op \frac{h}{2\pi}\op X^3 & 0 \\
        - \frac{\alpha'^2}{R_1^2} \op \frac{h}{2\pi}\op X^3   & R_2^2 + \frac{\alpha'^2}{R_1^2} \left[  
         \frac{h}{2\pi}\op  X^3 \right]^2& 0 \\
        0 & 0 & R_3^2
  \end{array}
  \right) ,
  \hspace{40pt}
  \check{\mathsf B}_{IJ} = 0 \,.
}
This background is known as a  twisted three-torus
\cite{Dasgupta:1999ss,Kachru:2002sk}, and the dual basis can be read-off from 
\eqref{basis_11} as $\{ \der\check\chi_1, \dx{2},\dx{3} \}$. 
The boundary condition for $\der\check\chi_1$ is determined via \eqref{neum_bc_a_23},
and together with the remaining directions we have
\eq{
  0 = \bigl(\der{\check\chi_1}\bigr)_{\rm tan} \,, \hspace{40pt}
  0 = \bigl(\dx{2}\bigr)_{\rm tan} \,, \hspace{40pt}
  0 = \bigl(\dx{3}\bigr)_{\rm tan} \,.
}
The dual background therefore contains a D0-brane. 
The residual 
$B$-field \eqref{dual_bres} can be determined as $\check B^{\rm res.} = \der X^1 \wedge \der \chi_1$,
and by performing the path integral over $X^1$ gives the condition 
\eq{
  \label{ex_3874544}
  \bigl[ \op \check\chi_1 - 2\pi \op a_1 \op \bigr]_{\partial\Sigma} \in 2\pi\op  \mathbb Z \,.
}
However, since we do not know how to quantize the theory in the presence of a
non-trivial $H$-flux we have no information about the momentum/winding numbers of the original 
coordinate $X^1$. Strictly speaking we should therefore set the right-hand side of 
\eqref{ex_3874544} to zero 
following \eqref{quant_dual_3}.
In summary, we see that the dual background is a twisted torus with a D0-brane, whose position 
is specified by the Wilson line $a_1$.


\subsubsection*{D2-brane along $X^1$\op--\op$X^2$}

As a second example, we consider a D2-brane along the directions $X^1$ and $X^2$ 
with a non-trivial open-string field strength $F_{12} = f={\rm const.}$ 
The boundary conditions \eqref{boundary_cond} then take the form
\eq{
  \label{ex_247}
  &0 = R_1^2\op \bigl(\dx{1}\bigr)_{\rm norm} + 2\pi\alpha'\op i \left( f + \tfrac{h}{4\pi^2} \op X^3 \right)
  \bigl(\dx{2}\bigr)_{\rm tan} \,, 
  \\[4pt]
  &0 = R_2^2\op \bigl(\dx{2}\bigr)_{\rm norm} - 2\pi\alpha'\op i \left( f + \tfrac{h}{4\pi^2} \op X^3 \right)
  \bigl(\dx{1}\bigr)_{\rm tan} \,, 
  \\[4pt]
  &0 = \bigl(\dx{3}\bigr)_{\rm tan} \,,
}
and for a T-duality along the direction $X^1$ the constraints  \eqref{constr1} and  \eqref{16_1}
are solved by
\eq{
  \arraycolsep2pt
  \begin{array}{lcl}
  a & = & \displaystyle  a_1 \op \dx{1} + 
  a_2 \op\dx{2} +f\op X^1 \op \dx{2} \,,
  \\[4pt]
  v_1 & =& \displaystyle  -2\pi\alpha' \op f\op \dx{2}\,,
  \\[4pt]
  \omega_1 &= &0 \,.
  \end{array}
  \hspace{40pt} a_1,a_2,f= {\rm const.}\,,
}
The T-dual metric and $B$-field are again determined from the general expressions 
\eqref{dual_g_007} and \eqref{dual_b_001} with \eqref{con_dual}, from which we find
\eq{
  \label{ex_42424}
  &
  \arraycolsep4pt
  \renewcommand{\arraystretch}{1.2}
\check{\mathsf G}_{IJ} = 
  \left( \begin{array}{ccc}
        \frac{\alpha'^2}{R_1^2} & - \frac{\alpha'^2}{R_1^2} \left[ 2\pi f+\frac{h}{2\pi}\op  X^3 \right]& 0 \\
        - \frac{\alpha'^2}{R_1^2} \left[ 2\pi f+\frac{h}{2\pi}\op  X^3 \right]  & R_2^2 + \frac{\alpha'^2}{R_1^2} \left[ 2\pi f+\frac{h}{2\pi}\op  X^3 \right]^2& 0 \\
        0 & 0 & R_3^2
  \end{array}
  \right) ,
  \\[4pt]
  &\check{\mathsf B}_{IJ} = 0 \,,
}
which again describes a twisted three-torus. Note however that here the gauge-invariant
open-string field strength $2\pi\alpha'\mathcal F_{12} = 2\pi \alpha' f + \frac{\alpha'}{2\pi}\op h X^3$ 
appears.
The dual basis is determined via 
\eqref{basis_11} and reads as before $\{ \der\check\chi_1, \dx{2},\dx{3} \}$, and 
from \eqref{neum_bc_a_23} we find the boundary conditions
\eq{
  \label{ex_bundary_24249}
  &0 = \bigl(\der{\check\chi_1}\bigr)_{\rm tan} \,,
  \\[4pt]
  &0 = \check{\mathsf G}_{2}{}^{1}\bigl(\der\check\chi_1\bigr)_{\rm norm}+ \check{\mathsf G}_{22}\bigl(\dx{2}\bigr)_{\rm norm} \,,  
  \\[4pt]
  &0 = \bigl(\dx{3}\bigr)_{\rm tan} \,.
}
These describe Dirichlet  conditions for $\check\chi_1$ and $X^3$ and a Neumann boundary condition
for $X^2$, and hence the dual backgrounds contains a D1-brane along the $X^2$-direction. 
The residual $B$-field \eqref{dual_bres} is determined 
as $\check B^{\rm res.} = \der X^1 \wedge (\der\chi_1-2\pi\alpha' f\op \dx{2})$, which 
via \eqref{exp_dual_2947} cancels the open-string gauge field $f\op X^1\op\dx{2}\subset a$
on the boundary. 
Performing then the path integral over $X^1$ gives 
\eq{
  \bigl[ \op \check\chi_1 - 2\pi \op a_1 \op \bigr]_{\partial\Sigma} =0 \,,
}
following the same reasoning leading to \eqref{quant_dual_3}.
The Wilson line $a_2 \op\dx{2}\subset a$ is 
untouched, so the dual open-string gauge field reads
\eq{
  \check{\mathsf a} = a_2 \op \dx{2} \,.
}
In summary, we find that the T-dual background is a twisted torus with a D1-brane with constant Wilson line
along the $X^2$-direction. 
We also note that when turning-off the $H$-flux and setting $h=0$, the metric \eqref{ex_42424} becomes
constant. The boundary conditions \eqref{ex_bundary_24249} then describe
a D1-brane at an angle in the $X^1$--$X^2$ torus, which reproduces the well-known CFT result.


\subsubsection*{D3-brane along $X^1$\op--\op$X^2$\op--\op$X^3$}
\label{page_d3_fw}

Let us also briefly discuss a D3-brane along all directions of the three-torus. 
Such a configuration does not satisfy the Freed-Witten anomaly cancellation condition \cite{Freed:1999vc}, 
which says that $H$ pulled-back to the D-brane has to vanish in cohomology. 
Nevertheless, we perform a T-duality transformation along the 
$X^1$-direction in order to gain insight on the dual background.

For simplicity, we consider a setting similar to the above-discussed D2-brane along $X^1$--$X^2$
but replace the Dirichlet boundary conditions along $X^3$ by Neumann boundary conditions
and set $a=0$. The dual background is then given by
\eq{
  \arraycolsep4pt
  \renewcommand{\arraystretch}{1.2}
  \check{\mathsf G}_{IJ} = 
  \left( \begin{array}{ccc}
        \frac{\alpha'^2}{R_1^2} & - \frac{\alpha'^2}{R_1^2} \op \frac{h}{2\pi}\op X^3 & 0 \\
        - \frac{\alpha'^2}{R_1^2} \op \frac{h}{2\pi}\op X^3   & R_2^2 + \frac{\alpha'^2}{R_1^2} \left[  
         \frac{h}{2\pi}\op  X^3 \right]^2& 0 \\
        0 & 0 & R_3^2
  \end{array}
  \right) ,
  \hspace{40pt}
  \check{\mathsf B}_{IJ} = 0 \,,
}
with boundary conditions
\eq{
  &0 = \bigl(\der{\check\chi_1}\bigr)_{\rm tan} \,,
  \\[4pt]
  &0 = \check{\mathsf G}_{2}{}^{1}\bigl(\der\check\chi_1\bigr)_{\rm norm}+ \check{\mathsf G}_{22}\bigl(\dx{2}\bigr)_{\rm norm} \,,  
  \\[4pt]
  &0 = \bigl(\dx{3}\bigr)_{\rm norm} \,.
}
These relations describe a D2-brane with Dirichlet boundary conditions along $\chi_1$ 
and Neumann conditions along $X^2$ and $X^3$. 
Since the original configuration is inconsistent,  this T-dual configuration 
has to be inconsistent as well. 
We come back to this point in section~\ref{sec_fwa}.


\subsubsection*{D0-brane}

We also want to illustrate T-duality transformations along Dirichlet directions. 
To do so, we first consider a D0-brane which is point-like on the three torus. 
The boundary conditions therefore read
\eq{
  0 = \bigl(\dx{1}\bigr)_{\rm tan} \,, \hspace{40pt}
  0 = \bigl(\dx{2}\bigr)_{\rm tan} \,, \hspace{40pt}
  0 = \bigl(\dx{3}\bigr)_{\rm tan} \,,
}
and the constraints \eqref{constr1} are solved by 
\eq{
  a = 0 \,, \hspace{40pt} v_1 = 0 \,, \hspace{40pt} \omega_1 = 0 \,.
}
The dual background is then given by 
\eqref{dual_g_007} and \eqref{dual_b_001} as
\eq{
  \arraycolsep4pt
  \renewcommand{\arraystretch}{1.2}
  \check{\mathsf G}_{IJ} = 
  \left( \begin{array}{ccc}
        \frac{\alpha'^2}{R_1^2} & - \frac{\alpha'^2}{R_1^2} \op \frac{h}{2\pi}\op X^3 & 0 \\
        - \frac{\alpha'^2}{R_1^2} \op \frac{h}{2\pi}\op X^3   & R_2^2 + \frac{\alpha'^2}{R_1^2} \left[  
         \frac{h}{2\pi}\op  X^3 \right]^2& 0 \\
        0 & 0 & R_3^2
  \end{array}
  \right) ,
  \hspace{40pt}
  \check{\mathsf B}_{IJ} = 0 \,,
}
which describes again a twisted three-torus.
The dual basis is determined via 
\eqref{basis_11} and reads  $\{ \der\check\chi_1, \dx{2},\dx{3} \}$, which 
satisfy the boundary conditions 
\eq{
  &0 = \check{\mathsf G}^{11}\bigl(\der{\check\chi_1}\bigr)_{\rm norm} + \check{\mathsf G}^1{}_2 \bigl(\dx{2}\bigr)_{\rm norm}  \,,
  \\[4pt]
  &0 = \bigl(\dx{2}\bigr)_{\rm tan}  \,,  
  \\[4pt]
  &0 = \bigl(\dx{3}\bigr)_{\rm tan}  \,.
}
These expressions describe a Neumann condition for the direction  $\check\chi_1$ and Dirichlet boundary conditions
for $X^2$ and $X^3$, and hence the dual background contains a D1-brane. 
The residual $B$-field \eqref{dual_bres} takes the form $\check B^{\rm res.}= \der X^1 \wedge \der\chi_1$, and
since $X^1$ satisfies Dirichlet boundary conditions $\der X^1$ is exact and we can compute
\eq{
  \label{ex_934792485}
  -\frac{i}{2\pi\alpha'}\int_{\Sigma} \check B^{\rm res.}  = 
  -\frac{i}{2\pi\alpha'}\int_{\partial\Sigma} 2\pi \alpha' \left[ \tfrac{X^1}{2\pi }\, \der \check\chi_1 \right].
}
We can therefore identify a constant Wilson line along the direction $\check\chi_1$ with the 
position of the D-brane along the original direction $X^1$
\eq{
  \check{\mathsf a} = \frac{X^1\rvert_{\partial\Sigma}}{2\pi}\, \der \check\chi_1  \,.
}
In summary, the T-dual background is a twisted three-torus with a D1-brane and 
a constant Wilson line corresponding to the position of D0-brane along the original direction $X^1$.


\subsubsection*{D1-brane along $X^2$}

As a second example for a T-duality transformation along a Dirichlet direction we 
consider a D1-brane along the $X^2$-direction. We choose a constant 
Wilson line for the D1-brane, and 
the boundary conditions read
\eq{
  0 = \bigl(\dx{1}\bigr)_{\rm tan} \,, \hspace{40pt}
  0 = \bigl(\dx{2}\bigr)_{\rm norm} \,, \hspace{40pt}
  0 = \bigl(\dx{3}\bigr)_{\rm tan} \,.
}
The constraints \eqref{constr1} are 
solved for instance by 
\eq{
  \arraycolsep2pt
  \begin{array}{lcl}
  a & = & \displaystyle  a_2 \op \dx{2} \,,
  \\[4pt]
  v_1 & =& \displaystyle  0\,,
  \\[4pt]
  \omega_1 &= &0 \,,
  \end{array}
  \hspace{80pt} a_2= {\rm const.}\,,
}
and the dual background is given again by 
\eq{
  \label{ex_9834691364}
  \arraycolsep4pt
  \renewcommand{\arraystretch}{1.2}
  \check{\mathsf G}_{IJ} = 
  \left( \begin{array}{ccc}
        \frac{\alpha'^2}{R_1^2} & - \frac{\alpha'^2}{R_1^2} \op \frac{h}{2\pi}\op X^3 & 0 \\
        - \frac{\alpha'^2}{R_1^2} \op \frac{h}{2\pi}\op X^3   & R_2^2 + \frac{\alpha'^2}{R_1^2} \left[  
         \frac{h}{2\pi}\op  X^3 \right]^2& 0 \\
        0 & 0 & R_3^2
  \end{array}
  \right) ,
  \hspace{40pt}
  \check{\mathsf B}_{IJ} = 0 \,,
}
with dual basis $\{ \der\check\chi_1, \dx{2},\dx{3} \}$. The boundary conditions  \eqref{dirc_bound_cond} 
are evaluated as
\eq{
  &0 = \check{\mathsf G}^{11}\bigl(\der{\check\chi_1}\bigr)_{\rm norm} + \check{\mathsf G}^1{}_2 \bigl(\dx{2}\bigr)_{\rm norm}  \,,
  \\[4pt]
  &0 = \check{\mathsf G}_2{}^{1}\bigl(\der{\check\chi_1}\bigr)_{\rm norm} + \check{\mathsf G}_{22} \bigl(\dx{2}\bigr)_{\rm norm}  \,  ,
  \\[4pt]
  &0 = \bigl(\dx{3}\bigr)_{\rm tan}  \,,
}
which describe a D2-brane along the directions $\check\chi_1$ and $X^2$. 
For the residual $B$-field a computation similar to \eqref{ex_934792485} applies, which 
leads to the following dual open-string gauge field
\eq{
  \check{\mathsf a} = \frac{X^1\rvert_{\partial\Sigma}}{2\pi }\, \der \check\chi_1  + a_2\op\dx{2} \,.
}


\subsubsection*{D2-brane along $X^2$\op--\op$X^3$}

For completeness, let us also consider a D2-brane along the directions $X^2$ and 
$X^3$ with vanishing open-string field strength. The analysis 
is very similar to the case of a D1-brane along $X^2$ which we just discussed. 
The dual background is given by \eqref{ex_9834691364}
with dual basis $\{ \der\check\chi_1, \dx{2},\dx{3} \}$. The boundary conditions  \eqref{dirc_bound_cond} 
are evaluated as
\eq{
  &0 = \check{\mathsf G}^{11}\bigl(\der{\check\chi_1}\bigr)_{\rm norm} + \check{\mathsf G}^1{}_2 \bigl(\dx{2}\bigr)_{\rm norm}  \,,
  \\[4pt]
  &0 = \check{\mathsf G}_2{}^{1}\bigl(\der{\check\chi_1}\bigr)_{\rm norm} + \check{\mathsf G}_{22} \bigl(\dx{2}\bigr)_{\rm norm}  \,  ,
  \\[4pt]
  &0 = \bigl(\dx{3}\bigr)_{\rm norm}  \,,
}
which describe a D3-brane along the twisted three-torus. 
Note that since the $H$-flux of this background vanishes, the Freed-Witten anomaly cancellation 
condition is satisfied.


\subsection{Two T-dualities}
\label{sec_2t}

In this section we consider two collective T-dualities for the three-torus with 
$H$-flux defined in \eqref{metr_bf}. For concreteness we 
always perform a collective duality transformation along the directions $X^1$ and 
$X^2$, which have the same boundary conditions. 
However, other combinations can be studied in a similar way.


\subsubsection*{D2-brane along $X^1$\op--\op$X^2$}

We start with a D2-brane along the directions $X^1$ and $X^2$ with a non-trivial open 
string field strength $F_{12}=f={\rm const.}$ The boundary conditions \eqref{boundary_cond}
then take the same form as in \eqref{ex_247}, namely
\eq{
  &0 = R_1^2\op \bigl(\dx{1}\bigr)_{\rm norm} + 2\pi\alpha'\op i \left( f + \tfrac{h}{4\pi^2} \op X^3 \right)
  \bigl(\dx{2}\bigr)_{\rm tan} \,, 
  \\[4pt]
  &0 = R_2^2\op \bigl(\dx{2}\bigr)_{\rm norm} - 2\pi\alpha'\op i \left( f + \tfrac{h}{4\pi^2} \op X^3 \right)
  \bigl(\dx{1}\bigr)_{\rm tan} \,, 
  \\[4pt]
  &0 = \bigl(\dx{3}\bigr)_{\rm tan} \,.
}
For a collective T-duality transformation along two directions the 
the constraints  \eqref{constr1} and \eqref{16_1}
are solved by
\eq{
  \arraycolsep2pt
  \begin{array}{lcl}
  a & = & \displaystyle  a_1 \op \dx{1} + 
  a_2 \op\dx{2} +\tfrac{1}{2} \op f \left( X^1 \op \dx{2} - X^2\op\dx{1} \right)\,,
  \\[4pt]
  v_1 & =& \displaystyle  -2\pi\alpha' \op f\op \dx{2}\,,
  \\[4pt]
  v_2 & =& \displaystyle  +2\pi\alpha' \op f\op \dx{1}\,,
  \\[4pt]
  \omega_1 &= & \displaystyle  -\hphantom{2}\pi\alpha' \op f\op \hphantom{\der}X^2 \,,
  \\[4pt]
  \omega_2 &= & \displaystyle  +\hphantom{2}\pi\alpha' \op f\op \hphantom{\der} X^1 \,,
  \end{array}
  \hspace{20pt} a_1,a_2,f= {\rm const.}\,,
}
and the T-dual metric and $B$-field are  determined from the general expressions 
\eqref{dual_g_007} and \eqref{dual_b_001} together with our convention \eqref{con_dual}. We find
\eq{
  \label{ex_tfold_24533}
  &
  \arraycolsep4pt
  \renewcommand{\arraystretch}{1.2}
\check{\mathsf G}_{IJ} = 
  \left( \begin{array}{ccc}
       \frac{\alpha'^2R_2^2}{R_1^2 R_2^2 + \left[ 2\pi\alpha' f+\frac{\alpha'}{2\pi}\op h X^3 \right]^2}  & 0 & 0 \\
     0  & \frac{\alpha'^2 R_1^2}{R_1^2 R_2^2 + \left[ 2\pi\alpha' f+\frac{\alpha'}{2\pi}\op h X^3 \right]^2}& 0 \\
        0 & 0 & R_3^2
  \end{array}
  \right) ,
  \\[4pt]
  &\check{\mathsf B}_{IJ} =  \left( \begin{array}{ccc}
       0 & \frac{-\alpha'^2 \left[ 2\pi\alpha' f+\frac{\alpha'}{2\pi}\op h X^3 \right]}{R_1^2 R_2^2 + \left[ 2\pi\alpha' f+\frac{\alpha'}{2\pi}\op h X^3 \right]^2}   & 0 \\
      \frac{+ \alpha'^2\left[ 2\pi\alpha' f+\frac{\alpha'}{2\pi}\op h X^3 \right]}{R_1^2 R_2^2 + \left[ 2\pi\alpha' f+\frac{\alpha'}{2\pi}\op h X^3 \right]^2}     & 0& 0 \\
        0 & 0 & 0
  \end{array}
  \right) ,
}
which are the metric and $B$-field of the T-fold background \cite{Hull:2004in}. Note that here 
again the gauge-invariant open-string field strength 
$2\pi\alpha'\mathcal F_{12} =  2\pi\alpha' f+\frac{\alpha'}{2\pi}\op h X^3 $ appears, and hence
the open-string sector has an effect on the T-dual closed-string background. 
The dual basis is determined via 
\eqref{basis_11} and reads  $\{ \der\check\chi_1, \der\check\chi_2,\dx{3} \}$, and from 
\eqref{neum_bc_a_23} we obtain the boundary conditions
\eq{
  0 = \bigl(\der{\check\chi_1}\bigr)_{\rm tan} \,, \hspace{40pt}
  0 = \bigl(\der\check\chi_2\bigr)_{\rm tan} \,, \hspace{40pt}
  0 = \bigl(\dx{3}\bigr)_{\rm tan} \,,
}
describing a D0-brane. 
The residual $B$-field \eqref{dual_bres}
is found as $\check B{}^{\rm res.} = \dx{1}\wedge \der\chi_1 + \dx{2}\wedge \der\chi_2 -2\pi\alpha' f \op \dx{1}\wedge \dx{2}$, which via the computation below \eqref{exp_dual_2947} leads to the 
 condition \eqref{quant_dual_3}
\eq{\label{quant_dual_ex2td_1}
  \bigl[ \op  \check\chi_{\alpha} - 2\pi \op a_{\alpha} \op \bigr]_{\partial\Sigma} =0 \,.
}


\subsubsection*{D3-brane along $X^1$\op--\op$X^2$\op--\op$X^3$}
\label{page_fw_3}

For later purposes let us also consider a D3-brane along the three-torus. Since 
here the Freed-Witten anomaly is not cancelled, this configuration is inconsistent. 
Nevertheless, applying a collective T-duality transformation along the 
directions $X^1$ and $X^2$ gives the T-fold background
\eqref{ex_tfold_24533}
with a D1-brane satisfying the boundary conditions
\eq{
  0 = \bigl(\der{\check\chi_1}\bigr)_{\rm tan} \,, \hspace{40pt}
  0 = \bigl(\der\check\chi_2\bigr)_{\rm tan} \,, \hspace{40pt}
  0 = \bigl(\dx{3}\bigr)_{\rm norm} \,.
}
Since the original configuration is not allowed by the Freed-Witten anomaly, this 
T-dual configuration is forbidden as well.
We come back to this point in section~\ref{sec_fwa}.


\subsubsection*{D0-brane}

We now turn to collective T-duality transformations along Dirichlet directions. 
For a D0-brane the boundary conditions take the form 
\eq{
  0 = \bigl(\dx{1}\bigr)_{\rm tan} \,, \hspace{40pt}
  0 = \bigl(\dx{2}\bigr)_{\rm tan} \,, \hspace{40pt}
  0 = \bigl(\dx{3}\bigr)_{\rm tan} \,,
}
and the constraints \eqref{constr1} are solved by 
\eq{
  a = 0 \,, \hspace{40pt} v_{1,2} = 0 \,, \hspace{40pt} \omega_{1,2} = 0 \,.
}
The dual background is then determined by the general expressions
\eqref{dual_g_007} and \eqref{dual_b_001} as
\eq{
  \label{ex_93846}
  &
  \arraycolsep4pt
  \renewcommand{\arraystretch}{1.2}
\check{\mathsf G}_{IJ} = 
  \left( \begin{array}{ccc}
       \frac{\alpha'^2R_2^2}{R_1^2 R_2^2 + \left[ \frac{\alpha'}{2\pi}\op h X^3 \right]^2}  & 0 & 0 \\
     0  & \frac{\alpha'^2R_1^2}{R_1^2 R_2^2 + \left[ \frac{\alpha'}{2\pi}\op h X^3 \right]^2}& 0 \\
        0 & 0 & R_3^2
  \end{array}
  \right) ,
  \\[4pt]
  &\check{\mathsf B}_{IJ} =  \left( \begin{array}{ccc}
       0 & \frac{- \alpha'^2\frac{\alpha'}{2\pi}\op h X^3}{R_1^2 R_2^2 + \left[ \frac{\alpha'}{2\pi}\op h X^3 \right]^2}   & 0 \\
      \frac{+ \alpha'^2\frac{\alpha'}{2\pi}\op h X^3 }{R_1^2 R_2^2 + \left[ \frac{\alpha'}{2\pi}\op h X^3 \right]^2}     & 0& 0 \\
        0 & 0 & 0
  \end{array}
  \right) ,
}
which is again that of a T-fold. Note that here the open-string gauge flux $F_{12}= f$ is absent, 
since the original D0-brane does not support an open-string gauge field.  The boundary conditions \eqref{dirc_bound_cond} 
lead to the following expressions
\eq{
  &0 = \check{\mathsf G}^{11}\bigl(\der{\check\chi_1}\bigr)_{\rm norm} 
  + i\, \check{\mathsf B}^{12} \bigl(\der{\check\chi_2}\bigr)_{\rm tan}  
  \,,
  \\[4pt]
  &0 = \check{\mathsf G}^{22}\bigl(\der{\check\chi_2}\bigr)_{\rm norm} 
  + i\, \check{\mathsf B}^{21} \bigl(\der{\check\chi_1}\bigr)_{\rm tan}  
  \,,
  \\[4pt]
  &0 = \bigl(\dx{3}\bigr)_{\rm tan}  \,,
}
which take the expected form of Neumann boundary conditions \eqref{boundary_cond}
for the dual coordinates.


\subsubsection*{D1-brane along $X^3$}

For completeness we also consider a D1-brane along the $X^3$-direction. 
This configuration is very similar to the case of a D0-brane which we just discussed, and 
a collective T-duality along the directions $X^1$ and $X^2$ gives the 
T-fold background \eqref{ex_93846}. 
The boundary conditions of the dual background 
describe a D3-brane
and are given by 
\eq{
  &0 = \check{\mathsf G}^{11}\bigl(\der{\check\chi_1}\bigr)_{\rm norm} 
  + i\, \check{\mathsf B}^{12} \bigl(\der{\check\chi_2}\bigr)_{\rm tan}  
  \,,
  \\[4pt]
  &0 = \check{\mathsf G}^{22}\bigl(\der{\check\chi_2}\bigr)_{\rm norm} 
  + i\, \check{\mathsf B}^{21} \bigl(\der{\check\chi_1}\bigr)_{\rm tan}  
  \,,
  \\[4pt]
  &0 = \bigl(\dx{3}\bigr)_{\rm norm}  \,.
}


\subsection{Three T-dualities}

We finally discuss a collective T-duality transformations for the 
three-torus along the directions $X^1$, $X^2$ and $X^3$. As we can 
see from  the second relation in \eqref{16_1}, in this 
case the $H$-flux has to vanish and we therefore set $h=0$ in \eqref{metr_bf}.


\subsubsection*{D3-brane along $X^1$\op--\op$X^2$\op--\op$X^3$}

Let us start with a D3-brane along all directions of the three-torus, and consider
an open-string field strength 
 $F_{12}=f={\rm const.}$ together with  $B=0$. The boundary conditions \eqref{boundary_cond}
 then read
\eq{
  &0 = R_1^2\op \bigl(\dx{1}\bigr)_{\rm norm} + 2\pi\alpha'\op i \op f \op
  \bigl(\dx{2}\bigr)_{\rm tan} \,, 
  \\[4pt]
  &0 = R_2^2\op \bigl(\dx{2}\bigr)_{\rm norm} - 2\pi\alpha'\op i \op f \op
  \bigl(\dx{1}\bigr)_{\rm tan} \,, 
  \\[4pt]
  &0 = R_3^2\op \bigl(\dx{3}\bigr)_{\rm norm} \,,
}
and the  constraints  \eqref{constr1} and \eqref{16_1}
are solved by
\eq{
  \arraycolsep2pt
  \begin{array}{lcl}
  a & = & \displaystyle  a_1 \op \dx{1} + 
  a_2 \op\dx{2}+ a_2 \op\dx{2} +\tfrac{1}{2} \op f \left( X^1 \op \dx{2} - X^2\op\dx{1} \right)\,,
  \\[4pt]
  v_1 & =& \displaystyle  -2\pi\alpha' \op f\op \dx{2}\,,
  \\[4pt]
  v_2 & =& \displaystyle  +2\pi\alpha' \op f\op \dx{1}\,,
  \\[4pt]
  v_3 & =& \displaystyle 0\,,
  \\[4pt]  \omega_1 &= & \displaystyle  -\hphantom{2}\pi\alpha' \op f\op \hphantom{\der}X^2 \,,
  \\[4pt]
  \omega_2 &= & \displaystyle  +\hphantom{2}\pi\alpha' \op f\op \hphantom{\der} X^1 \,,
  \\[4pt]
  \omega_3 &= & \displaystyle  0\,.
  \end{array}
  \hspace{-40pt} a_1,a_2,a_3,f= {\rm const.}\,,
}
The T-dual metric and $B$-field are  determined from the general expressions 
\eqref{dual_g_007} and \eqref{dual_b_001} for which we find
\eq{
  &
  \arraycolsep4pt
  \renewcommand{\arraystretch}{1.2}
\check{\mathsf G}_{IJ} = 
  \left( \begin{array}{ccc}
       \frac{\alpha'^2R_2^2}{R_1^2 R_2^2 + \left[ 2\pi\alpha' f\right]^2}  & 0 & 0 \\
     0  & \frac{\alpha'^2R_1^2}{R_1^2 R_2^2 + \left[ 2\pi\alpha' f \right]^2}& 0 \\
        0 & 0 & \frac{\alpha'^2}{R_3^2}
  \end{array}
  \right) ,
  \\[4pt]
  &\check{\mathsf B}_{IJ} =  \left( \begin{array}{ccc}
       0 & \frac{-  2\pi\alpha'^3 f}{R_1^2 R_2^2 + \left[ 2\pi\alpha' f\right]^2}   & 0 \\
      \frac{+  2\pi\alpha'^3 f}{R_1^2 R_2^2 + \left[ 2\pi\alpha' f\right]^2}     & 0& 0 \\
        0 & 0 & 0
  \end{array}
  \right) ,
}
and we see again that the open-string gauge flux $f$ enters the T-dual closed-string background.
The dual basis
$\{ \der\check\chi_1, \der\check\chi_2,\der\check\chi_3 \}$ is subject to  the boundary conditions
describing  a D0-brane
\eq{
  0 = \bigl(\der{\check\chi_1}\bigr)_{\rm tan} \,, \hspace{40pt}
  0 = \bigl(\der\check\chi_2\bigr)_{\rm tan} \,, \hspace{40pt}
  0 = \bigl(\der\check\chi_{3}\bigr)_{\rm tan} \,,
}
where the location of each dual coordinate on the boundary is given by 
the open-string gauge field as already shown in \eqref{quant_dual_ex2td_1}.


\subsubsection*{D0-brane}

Finally, for a D0-brane there is no open-string field strength and hence the 
dual metric and $B$-field after a collective T-duality transformation along all 
three directions read
\eq{
  \arraycolsep4pt
  \renewcommand{\arraystretch}{1.2}
  \check{\mathsf G}_{IJ} = 
  \left( \begin{array}{ccc}
        \frac{\alpha'^2}{R_1^2} & 0 & 0 \\
        0 & \frac{\alpha'^2}{R_2^2}& 0 \\
        0 & 0 & \frac{\alpha'^2}{R_3^2}
  \end{array}
  \right) ,
  \hspace{60pt}
  \check{\mathsf B}_{IJ} = 0 \,.
}
The dual coordinates are subject to the boundary conditions
\eq{
  0 = \bigl(\der{\check\chi_1}\bigr)_{\rm norm} \,, \hspace{40pt}
  0 = \bigl(\der\check\chi_2\bigr)_{\rm norm} \,, \hspace{40pt}
  0 = \bigl(\der\check\chi_{3}\bigr)_{\rm norm} \,,
}
which describe a D3-brane. The  open-string gauge field 
is characterized by the position of the original D0-brane
as in \eqref{dir_324545}.


\section{Freed-Witten anomaly and boundary conditions}
\label{sec_discussion}

We now discuss the results obtained in section~\ref{sec_examples}. 
We first briefly review the Freed-Witten anomaly cancellation condition, 
and then study the global properties of the open-string boundary conditions.


\subsection{Freed-Witten anomaly}
\label{sec_fwa}

It is known that D-branes in backgrounds with non-vanishing $H$-flux are subject to 
the Freed-Witten anomaly cancellation condition \cite{Freed:1999vc}. In particular, the restriction of
the field-strength $H=\der B$ to the D-brane has to vanish (in cohomology). 
Denoting the cycle  wrapped by the D-brane by $\Gamma$
and its Poincar\'e dual by $[\Gamma]$,
this condition can be expressed as
\eq{
  H\wedge [\op\Gamma\op] = 0\,.
}
For backgrounds with geometric $F$-flux and non-geometric $Q$- and $R$-fluxes
the generalization of this condition has been discussed 
for instance in 
\cite{Camara:2005dc,Villadoro:2006ia,LoaizaBrito:2006se,Aldazabal:2008zza,Aldazabal:2011yz,Blumenhagen:2015kja}.
Here one finds the  expression 
\eq{
  \label{fw_002}
  \bigl( \der- H\wedge\: - F\circ\: - Q\bullet\: - R\,\llcorner \bigr) \op [\op \Gamma\op]=0\,,
}
where the various fluxes are interpreted as operators acting in $[\op\Gamma\op]$. Using the 
contraction with a vector field $\iota_i\equiv \iota_{\partial_i}$, in a coordinate basis they act as
\eq{
  \label{flux_ops}
  \arraycolsep1pt
  \begin{array}{l@{\hspace{4pt}}c@{\hspace{3pt}}cc@{\hspace{5pt}}cccccl}
  H \,\wedge &=&  \frac{1}{3!} & H_{ijk} & \der X^i &\wedge& \der X^j  &\wedge& \der X^k & \,, \\[6pt]
  F \,\circ &=&  \frac{1}{2!} & F^k{}_{ij} &  \der X^i  &\wedge& \der X^j & \wedge & \iota_k & \,,   \\[6pt]
  Q \,\bullet &=&  \frac{1}{2!} & Q_i{}^{jk} & \der X^i &\wedge& \iota_j  &\wedge& \iota_k & \,,   \\[6pt]  
  R \,\llcorner &=& \frac{1}{3!} & R^{ijk} & \iota_i &\wedge& \iota_j &\wedge& \iota_k & \,.   
  \end{array}
}
Let us now discuss this condition for the examples studied in section~\ref{sec_examples}:
\begin{itemize}

\item For the three-torus with $H$-flux we mentioned already on page~\pageref{page_d3_fw} that 
a D3-brane is forbidden by the Freed-Witten anomaly. And indeed, 
since $[\Gamma_{\rm D3}]$ is a point on $\mathbb T^3$ we see that
in this case \raisebox{0pt}[0pt][0pt]{$H\overset{!}{=}0$}.

\item For the twisted torus we can determine the geometric flux $F^k{}_{ij}$ as the structure 
constants of the vielbein one-forms under the exterior derivative. We
see that for the examples in section~\ref{sec:onetd} only $F^1{}_{23}$ is non-vanishing, 
and hence \eqref{fw_002} implies that on a twisted $\mathbb T^3$
a D2-brane along the directions $X^2$ and $X^3$ is forbidden. 
This is in agreement with our conclusion on page~\pageref{page_d3_fw}.

\item For the T-fold backgrounds obtained in section~\ref{sec_2t} 
we can determine the non-geometric $Q$-flux via \eqref{def_qr}. Here we find
that the only non-vanishing component is $Q_3{}^{12}$, and hence 
\eqref{fw_002} implies that on a T-fold a D1-brane along the $X^3$-direction is not
allowed. This is again in agreement with our findings on page~\pageref{page_fw_3}.

\end{itemize}


\subsection{Boundary conditions}

Next, we consider  the global behavior  of the open-string boundary conditions
for the backgrounds studied in the last section. To do so we first 
briefly recall how the examples of section~\ref{sec_examples} can be 
interpreted as torus fibrations over a circle, and then 
turn to the global properties of the boundary conditions.


\subsubsection*{Torus fibrations}

We note that  the three-torus with $H$-flux, the twisted three-torus and the T-fold background can 
all be realized as $\mathbb T^2$-fibrations over a circle. 
In particular, for the examples studied in section~\ref{sec_examples} we can express 
the metric and $B$-field as
\eq{
  \renewcommand{\arraystretch}{1.1}
  \arraycolsep3pt
  G_{ij} = \left( \begin{array}{cc}
  G_{\mathsf i\mathsf j}(X^3) & 0 \\ 0 & R_3^2 \end{array}\right),
  \hspace{50pt}
  B_{ij} = \left( \begin{array}{cc}
  B_{\mathsf i\mathsf j}(X^3) & 0 \\ 0 & 0 \end{array}\right),
}
with $i,j=1,2,3$ and $\mathsf i,\mathsf j = 1,2$ labelling the fiber directions. 
These fibrations are globally well-defined through gluing local charts with 
$O(D,D;\mathbb Z)$ transformations, which include gauge transformations, diffeomorphisms 
and so-called $\beta$-transformations. 
This can be made precise by  defining a generalized metric $\mathcal H$ which contains the metric and $B$-field
as
\eq{
  \arraycolsep4pt
  \mathcal H = \left( \begin{array}{cc}
   \frac{1}{\alpha'}\left( G - B\op G^{-1}B\right)& +B\op G^{-1}  \\[4pt]  - G^{-1} B & \alpha' G^{-1}\end{array} \right)
   ,
}
for which we can explicitly check that he examples of section~\ref{sec_examples} satisfy
\eq{
   \mathcal H \bigl( X^3+2\pi \bigr)= \mathcal O^{-T} \op\mathcal H (X^3) \:\mathcal O^{-1}
   \,.
}
Here, $\mathcal O\in O(2,2;\mathbb Z)\subset O(3,3;\mathbb Z)$ are transformations which take the form
\eq{
  \label{bc_002}
  \begin{array}{l@{\hspace{30pt}}l@{\hspace{30pt}}l}
  \mbox{$\mathbb T^3$ with $H$-flux:} & 
  \arraycolsep5pt
  \displaystyle \mathcal O_{\mathsf B}= \left( \begin{array}{cc} \mathds 1 & 0 \\ \mathsf B & \mathds 1 \end{array}\right)
  \,,
  &
  \resizebox{80pt}{!}{$
  \arraycolsep4pt
  \displaystyle \mathsf B = \left( \begin{array}{ccc} 0 & +h & 0 \\ -h & 0  & 0 \\ 0 & 0 & 0 \end{array}\right)
  $},
  \\[20pt]
  \mbox{twisted $\mathbb T^3$:} & 
  \arraycolsep2pt
  \displaystyle \mathcal O_{\mathsf A}=  \left( \begin{array}{cc} \mathsf A^{-1} & 0 \\ 0 & \mathsf A^{T} \end{array}\right)
  \,,
  &
  \resizebox{80pt}{!}{$
  \arraycolsep4pt
  \displaystyle  \mathsf A =  \left( \begin{array}{ccc} 1 & -h & 0 \\ 0 & 1 & 0 \\ 0 & 0 & 1\end{array}\right)
  $},
  \\[20pt]
  \mbox{T-fold:} & 
  \arraycolsep5pt
  \displaystyle \mathcal O_{\mathsf \beta}=  \left( \begin{array}{cc} \mathds 1 & \beta \\ 0 & \mathds 1 \end{array}\right) 
  \,,
  &
  \resizebox{80pt}{!}{$
  \arraycolsep4pt
  \displaystyle   \beta =  \left( \begin{array}{ccc} 0 & +h & 0 \\ -h & 0 & 0 \\ 0 & 0 & 0\end{array}\right)
  $},
  \end{array}
}
and which correspond to gauge transformations, diffeomorphisms and $\beta$-trans\-for\-mations, 
respectively.


\subsubsection*{Boundary conditions}

Let us now turn to the boundary conditions. 
Using $2D$-dimensional matrix notation we can
express the Dirichlet and Neumann conditions shown in \eqref{boundary_cond} in the following way
\eq{
  \label{bc_001}
  \renewcommand{\arraystretch}{1.1}
  \arraycolsep4pt
  \left( \begin{array}{@{\hspace{2pt}}c@{\hspace{2pt}}}\rm D \\ \rm N \end{array}\right) 
   = \left( \begin{array}{cc}
  \alpha' & 0 \\
  2\pi \alpha' \mathcal F & G 
  \end{array}\right)
  \left( \begin{array}{@{\hspace{2pt}}c@{\hspace{1pt}}l@{\hspace{2pt}}}
  i &  \bigl(\der X\bigr)_{\rm tan} \\
  & \bigl(\der X\bigr)_{\rm norm}
  \end{array}\right),
}
where the restriction of $G$ and $\mathcal F$ to the boundary $\partial\Sigma$ is understood. 
The dilaton can be studied separately and we come back to it below. 
A particular D-brane configuration is then specified by a projection operator $\Pi$ acting 
on \eqref{bc_001}, which takes the general form
\eq{
  \label{bc_004}
  \Pi = \left( \begin{array}{cc} \Delta & 0 \\ 0 & \mathds 1 - \Delta \end{array}\right)
  ,
  \hspace{50pt}
  \Delta^2 = \Delta\,.
}
For instance, a D1-brane along the $X^1$-direction is characterized by the 
$D\times D$ matrix $\Delta = {\rm diag}\op(0,1,\ldots, 1)$.

We now want to determine how the boundary conditions \eqref{bc_001} 
of the three-torus with $H$-flux, twisted three-torus and the T-fold 
background behave under $X^3\to X^3+ 2\pi$.
For the coordinates we find that under $O(D,D;\mathbb Z)$ transformations 
we have the following general behavior fiber-wise\,\footnote{
For a general transformation of the form $\mathcal O = 
\raisebox{1pt}{\resizebox{!}{10pt}{$\left( \arraycolsep2pt \begin{array}{cc} a & b \\ c & d \end{array}\right)$}}\in
O(D,D;\mathbb Z)$ the matrix $\Omega$ can be determined as
$\Omega=\op\raisebox{1pt}{\resizebox{!}{10pt}{$\left( \arraycolsep2pt \begin{array}{cc}
 a + 2\pi \op b\op \mathcal F&  \frac{1}{\alpha'} \op b\,  G \\ \frac{1}{\alpha'}\op b\, G &  a + 2\pi \op b\op \mathcal F \end{array}\right)$}}$,
with $G$ the metric and $\mathcal F$ the gauge-invariant field-strength.
}
\eq{
  \label{bc_003}
\left( \begin{array}{@{\hspace{2pt}}c@{\hspace{1pt}}l@{\hspace{2pt}}}
  i &  \bigl(\der X\bigr)_{\rm tan} \\
  & \bigl(\der X\bigr)_{\rm norm}
  \end{array}\right)
   \quad   \xrightarrow{\hspace{10pt}\mathcal O\hspace{10pt}} \quad
\left( \begin{array}{@{\hspace{2pt}}c@{\hspace{1pt}}l@{\hspace{2pt}}}
  i &  \bigl(\der \tilde X\bigr)_{\rm tan} \\
  & \bigl(\der \tilde X\bigr)_{\rm norm}
  \end{array}\right)
=
  \Omega
    \left( \begin{array}{@{\hspace{2pt}}c@{\hspace{1pt}}l@{\hspace{2pt}}}
  i &  \bigl(\der X\bigr)_{\rm tan} \\
  & \bigl(\der X\bigr)_{\rm norm}
  \end{array}\right),
}
where the $2D\times 2D$ matrix $\Omega$ for each of the cases takes the  form
\eq{
  \arraycolsep2pt
  \begin{array}{l@{\hspace{30pt}}lcl}
  \mbox{$\mathbb T^3$ with $H$-flux:} &
  \Omega_{\mathsf B} &=& 
  \arraycolsep4pt
   \displaystyle      \left( \begin{array}{cc}
  \mathds 1 & 0 \\[2pt] 0 & \mathds 1 \end{array}\right) ,
  \\[20pt]
  \mbox{twisted $\mathbb T^3$:} &
  \Omega_{\mathsf A} &=& 
  \arraycolsep4pt
   \displaystyle      \left( \begin{array}{cc}
  \mathsf A^{-1} & 0 \\[2pt] 0 & \mathsf A^{-1} \end{array}\right) ,
  \\[20pt]
  \mbox{T-fold:} &
  \Omega_{\beta} &=& 
  \arraycolsep4pt
   \displaystyle      \left( \begin{array}{cc}
  \mathds 1 + 2\pi\op\beta\op \mathcal F & \frac{1}{\alpha'} \op\beta\op G \\[2pt] \frac{1}{\alpha'} \op\beta\op G 
  & \mathds 1 + 2\pi\op\beta\op \mathcal F \end{array}\right) .
 \end{array}
}
The matrices $\mathsf A$ and $\beta$ have been defined in \eqref{bc_002}, and 
we note that for the case of the T-fold 
the normal and tangential part of $\dx{i}$ are mixed under the 
$O(D,D;\mathbb Z)$ transformation.
Using these relations, for the examples of section~\ref{sec_examples} we then find 
\eq{
  \label{trans_bc_03}
  \renewcommand{\arraystretch}{1.1}
  \arraycolsep4pt
  \left( \begin{array}{@{\hspace{2pt}}c@{\hspace{2pt}}}\rm D \\ \rm N \end{array}
  \right)_{\raisebox{2.5pt}{\scriptsize$X^3+2\pi$}} 
  & = \hphantom{\mathcal O_{\star} }\left( \begin{array}{cc}
  \alpha' & 0 \\
  2\pi \alpha' \mathcal F & G 
  \end{array}\right)_{\raisebox{2.5pt}{\scriptsize$X^3+2\pi$}} 
  \hspace{4pt}
  \left( \begin{array}{@{\hspace{2pt}}c@{\hspace{1pt}}l@{\hspace{2pt}}}
  i &  \bigl(\der \tilde X\bigr)_{\rm tan} \\
  & \bigl(\der \tilde X\bigr)_{\rm norm}
  \end{array}\right)
  \\[6pt]
  &=\mathcal O_{\star} 
  \left( \begin{array}{cc}
  \alpha' & 0 \\
  2\pi \alpha' \mathcal F & G 
  \end{array}\right)_{\raisebox{2.5pt}{\scriptsize$X^3$}} \,\Omega^{-1}_{\star}
  \left( \begin{array}{@{\hspace{2pt}}c@{\hspace{1pt}}l@{\hspace{2pt}}}
  i &  \bigl(\der \tilde X\bigr)_{\rm tan} \\
  & \bigl(\der \tilde X\bigr)_{\rm norm}
  \end{array}\right)
  \\[6pt]
  &=\mathcal O_{\star} 
  \renewcommand{\arraystretch}{1.1}
  \arraycolsep4pt
  \left( \begin{array}{@{\hspace{2pt}}c@{\hspace{2pt}}}\rm D \\ \rm N \end{array}
  \right)_{\raisebox{2.5pt}{\scriptsize$X^3$}}  \,,
}
where the subscript $\star=(\mathsf B,\mathsf A,\beta)$ corresponds to  the three-torus with $H$-flux, 
the twisted $\mathbb T^3$ and the T-fold. 
The coordinates $\der \tilde X^i$ in one patch are related to $\der X^i$ in another patch 
via \eqref{bc_003}, and 
we emphasize that these relations are to be evaluated on the boundary. 
We then see that the  boundary conditions are globally-well defined using, respectively,
gauge transformations, diffeomorphisms and $\beta$-transformations.


\subsubsection*{Dilaton}

In the expression \eqref{bc_001} for the open-string boundary conditions we have omitted 
the dilaton. This contribution can be discussed separately, and we first determine
using \eqref{dilaton}
\eq{
  \arraycolsep2pt
  \begin{array}{l@{\hspace{30pt}}lcl}
  \mbox{$\mathbb T^3$ with $H$-flux:} &
  \displaystyle \phi &=& \displaystyle \phi_0 \,,
  \\[20pt]
  \mbox{twisted $\mathbb T^3$:} &
  \displaystyle \phi &=& 
  \displaystyle \phi_0-\log\left[\frac{R_1}{\sqrt{\alpha'}} \right],
  \\[20pt]
  \mbox{T-fold:} &
  \displaystyle \phi &=& 
  \displaystyle \phi_0 - \frac{1}{2} \log\left[ \frac{R_1^2\op R_2^2}{\alpha'^2} + \left(2\pi f + \tfrac{h}{2\pi} X^3\right)^2 \right],
 \end{array}
}
where for the T-fold we included the open-string field strength $f={\rm const.}$ which 
in some examples vanishes. 
We now consider each of these cases separately:
\begin{itemize}

\item For the three torus with $H$-flux, a gauge transformation leaves the metric invariant,
and hence the combination $e^{-2\phi}\sqrt{\det G}$ is invariant under the action of $\mathcal O_{\mathsf B}$. 
Furthermore, since the dilaton is constant it does not change under $X^3\to X^3+2\pi$ and 
so the contribution to the corresponding boundary conditions is well-defined. 

\item For the twisted three-torus  $e^{-2\phi}\sqrt{\det G}$ is invariant 
under diffeomorphisms $\mathcal O_{\mathsf A}$ and the dilaton is constant, so the 
contribution to the boundary conditions is again well-defined.

\item For the T-fold on the other hand, the dilaton is not constant and transforms under 
$\beta$-transformations.
In particular, by requiring  $e^{-2\phi}\sqrt{\det G}$ to be invariant we can determine
\eq{
  \phi(X^3+2\pi) = \mathcal O_{\beta}\bigl[ \phi(X^3)\bigr] \,,
}
where the action of $\mathcal O_{\beta}$ is 
understood in an abstract way and not as a matrix multiplication.
We therefore see that the dilaton is well-defined under $X^3\to X^3+ 2\pi$
using a $\beta$-transformation, and
hence also the contribution to the boundary conditions is well-defined.

\end{itemize}


\subsubsection*{Projection}

So far we have studied how \eqref{bc_001} behaves under $X^3\to X^3+2\pi$ for the 
examples of section~\ref{sec_examples}.
We now want to discuss how the projection operator \eqref{bc_004} is implemented on the boundary 
conditions. To do so, we again proceed by discussing the examples:
\begin{itemize}

\item For the three-torus with $H$-flux the behavior under 
$X^3\to X^3+2\pi$ is captured by \eqref{trans_bc_03}, provided that 
first the $\mathcal O_{\mathsf B}\in O(3,3;\mathbb Z)$ transformation is performed and 
after that the projection \eqref{bc_004}. In particular, we have
\eq{
  \label{bc_294724}
  \renewcommand{\arraystretch}{1.1}
  \arraycolsep4pt
  \Pi \left[ 
  \left( \begin{array}{@{\hspace{2pt}}c@{\hspace{2pt}}}\rm D \\ \rm N \end{array}
  \right)_{\raisebox{2.5pt}{\scriptsize$X^3+2\pi$}} 
  \right]
  = \Pi \left[ \mathcal O_{\mathsf B} 
  \left( \begin{array}{@{\hspace{2pt}}c@{\hspace{2pt}}}\rm D \\ \rm N \end{array}
  \right)_{\raisebox{2.5pt}{\scriptsize$X^3$}}  
  \right] .
}
One quickly sees for instance  from the (NN) case 
that if we perform the $\mathcal O_{\mathsf B}$ transformation on the  projected boundary 
conditions we do not reproduce the expected result from \eqref{trans_bc_03}.

\item For the twisted three-torus a similar analysis can be made. We verified explicitly that 
a projection similar to \eqref{bc_294724} produces the expected behavior of the 
boundary conditions from  \eqref{trans_bc_03}, 
and that performing the $\mathcal O_{\mathsf A}$ transformation on the projected 
boundary conditions does not match with the explicit computation.

\item Finally, for the T-fold the condition \eqref{bc_294724} similarly applies. This
means in particular, that the type of D-brane does not change 
under the identification $X^3\to X^3+ 2\pi$. 
The boundary conditions are therefore well-defined.

\end{itemize}


\clearpage
\section{Summary and conclusion}
\label{sec_sum_con}

In this paper we have studied T-duality transformations for open-string backgrounds via Buscher's 
procedure.  We illustrated this formalism with the example of the three-torus with 
$H$-flux and its T-dual configurations, and we analyzed global properties 
of the open-string   boundary conditions for these backgrounds. 
More concretely:
\begin{itemize}

\item T-duality transformations for open strings via Buscher's procedure have 
been discussed before in the literature \cite{Alvarez:1996up,Dorn:1996an,Forste:1996hy}.
Here we  extended these analyses and worked-out missing details: we took into account non-trivial world-sheet 
topologies, we included T-duality along directions with Dirichlet boundary conditions, and 
we allowed for collective T-duality transformations along multiple directions.  

We find that -- as expected  -- also for curved backgrounds  Neumann and Dirichlet 
boundary conditions are interchanged under T-duality, and that a constant open-string 
Wilson line along a Neumann direction 
shifts the position of the D-brane in the T-dual Dirichlet direction and vice versa.

\item In section~\ref{sec_examples} we illustrated the above formalism through various examples 
for the three-torus with $H$-flux. We  obtained D-brane configurations on the 
twisted three-torus and on the T-fold, and we saw that an open-string gauge-flux  
affects  the closed-string sector of the T-dual theory.

\item In section~\ref{sec_discussion} we discussed the results of section~\ref{sec_examples}.
After briefly reviewing the Freed-Witten anomaly cancellation condition, we showed that 
D-brane boundary conditions for the three-torus with $H$-flux, the twisted three-torus and for the 
T-fold are globally well-defined using, respectively, gauge transformations, diffeomorphisms and 
$\beta$-transformations. 

Since $\beta$-transformations mix the tangential and normal part of $\der X^i$ on the boundary, 
naively one might have thought that D$p$-branes on the T-fold  can change their dimensionality 
under $X^3\to X^3+2\pi$. However, we show that this is not true due to the 
mixing between the metric and $B$-field under $\beta$-transformations. 
Our findings furthermore agree with results obtained in doubled geometry 
in \cite{Hull:2004in,Lawrence:2006ma,Albertsson:2008gq}.

\end{itemize}

An interesting next step is to extend our formalism to non-abelian T-duality transformations
\cite{
delaOssa:1992vci,
Giveon:1993ai,
Alvarez:1993qi,
Sfetsos:1994vz,
Alvarez:1994np,
Lozano:1995jx
}. 
We have already included the possibility of a non-abelian isometry algebra for the gauging 
procedure and for integrating-out the gauge fields, however, the change of 
basis \eqref{basis_11} is singular in certain non-abelian cases.  
One approach to avoid this problem is to find a different change of basis which is non-singular, 
and we hope to come back to this question in the future.


\vskip4em
\subsection*{Acknowledgements}

We would like to thank 
R.~Blumenhagen,
S.~Krippendorf,
C.~Mayrhofer
and  
\linebreak R.~Szabo 
for helpful discussions. 
The work of F.C.-T. was  funded by the CONICYT scholarship 
72160340  from the Government of Chile.
The work of D.L. is supported by the ERC Advanced Grant ``Strings and Gravity'' (Grant No. 320045) and the Excellence Cluster Universe. He also is grateful to the CERN theory department for its hospitality, when part of this work was performed.


\vskip3em

\bibliography{references}

\providecommand{\href}[2]{#2}\begingroup\raggedright\begin{thebibliography}{10}

\bibitem{Hellerman:2002ax}
S.~Hellerman, J.~McGreevy, and B.~Williams, ``{Geometric constructions of
  nongeometric string theories},'' {\em JHEP} {\bf 01} (2004) 024,
  \href{http://xxx.lanl.gov/abs/hep-th/0208174}{{\tt hep-th/0208174}}.

\bibitem{Hull:2004in}
C.~M. Hull, ``{A Geometry for non-geometric string backgrounds},'' {\em JHEP}
  {\bf 10} (2005) 065, \href{http://xxx.lanl.gov/abs/hep-th/0406102}{{\tt
  hep-th/0406102}}.

\bibitem{Dasgupta:1999ss}
K.~Dasgupta, G.~Rajesh, and S.~Sethi, ``{M theory, orientifolds and G -
  flux},'' {\em JHEP} {\bf 08} (1999) 023,
  \href{http://xxx.lanl.gov/abs/hep-th/9908088}{{\tt hep-th/9908088}}.

\bibitem{Kachru:2002sk}
S.~Kachru, M.~B. Schulz, P.~K. Tripathy, and S.~P. Trivedi, ``{New
  supersymmetric string compactifications},'' {\em JHEP} {\bf 03} (2003) 061,
  \href{http://xxx.lanl.gov/abs/hep-th/0211182}{{\tt hep-th/0211182}}.

\bibitem{Shelton:2005cf}
J.~Shelton, W.~Taylor, and B.~Wecht, ``{Nongeometric flux compactifications},''
  {\em JHEP} {\bf 10} (2005) 085,
  \href{http://xxx.lanl.gov/abs/hep-th/0508133}{{\tt hep-th/0508133}}.

\bibitem{Shelton:2006fd}
J.~Shelton, W.~Taylor, and B.~Wecht, ``{Generalized Flux Vacua},'' {\em JHEP}
  {\bf 02} (2007) 095, \href{http://xxx.lanl.gov/abs/hep-th/0607015}{{\tt
  hep-th/0607015}}.

\bibitem{Mathai:2004qq}
V.~Mathai and J.~M. Rosenberg, ``{T duality for torus bundles with H fluxes via
  noncommutative topology},'' {\em Commun. Math. Phys.} {\bf 253} (2004)
  705--721, \href{http://xxx.lanl.gov/abs/hep-th/0401168}{{\tt
  hep-th/0401168}}.

\bibitem{Mathai:2004qc}
V.~Mathai and J.~M. Rosenberg, ``{On Mysteriously missing T-duals, H-flux and
  the T-duality group},'' in {\em {Differential geometry and physics.
  Proceedings, 23rd International Conference, Tianjin, China, August 20-26,
  2005}}, pp.~350--358, 2004.
\newblock \href{http://xxx.lanl.gov/abs/hep-th/0409073}{{\tt hep-th/0409073}}.

\bibitem{Lust:2010iy}
D.~L{\"u}st, ``{T-duality and closed string non-commutative (doubled)
  geometry},'' {\em JHEP} {\bf 12} (2010) 084,
  \href{http://xxx.lanl.gov/abs/1010.1361}{{\tt 1010.1361}}.

\bibitem{Condeescu:2012sp}
C.~Condeescu, I.~Florakis, and D.~L{\"u}st, ``{Asymmetric Orbifolds,
  Non-Geometric Fluxes and Non-Commutativity in Closed String Theory},'' {\em
  JHEP} {\bf 04} (2012) 121, \href{http://xxx.lanl.gov/abs/1202.6366}{{\tt
  1202.6366}}.

\bibitem{Andriot:2012an}
D.~Andriot, O.~Hohm, M.~Larfors, D.~L{\"u}st, and P.~Patalong, ``{Non-Geometric
  Fluxes in Supergravity and Double Field Theory},'' {\em Fortsch. Phys.} {\bf
  60} (2012) 1150--1186, \href{http://xxx.lanl.gov/abs/1204.1979}{{\tt
  1204.1979}}.

\bibitem{Andriot:2012vb}
D.~Andriot, M.~Larfors, D.~L{\"u}st, and P.~Patalong, ``{(Non-)commutative
  closed string on T-dual toroidal backgrounds},'' {\em JHEP} {\bf 06} (2013)
  021, \href{http://xxx.lanl.gov/abs/1211.6437}{{\tt 1211.6437}}.

\bibitem{Blair:2014kla}
C.~D.~A. Blair, ``{Non-commutativity and non-associativity of the doubled
  string in non-geometric backgrounds},'' {\em JHEP} {\bf 06} (2015) 091,
  \href{http://xxx.lanl.gov/abs/1405.2283}{{\tt 1405.2283}}.

\bibitem{Bouwknegt:2004ap}
P.~Bouwknegt, K.~Hannabuss, and V.~Mathai, ``{Nonassociative tori and
  applications to T-duality},'' {\em Commun. Math. Phys.} {\bf 264} (2006)
  41--69, \href{http://xxx.lanl.gov/abs/hep-th/0412092}{{\tt hep-th/0412092}}.

\bibitem{Ellwood:2006my}
I.~Ellwood and A.~Hashimoto, ``{Effective descriptions of branes on
  non-geometric tori},'' {\em JHEP} {\bf 12} (2006) 025,
  \href{http://xxx.lanl.gov/abs/hep-th/0607135}{{\tt hep-th/0607135}}.

\bibitem{Blumenhagen:2010hj}
R.~Blumenhagen and E.~Plauschinn, ``{Nonassociative Gravity in String
  Theory?},'' {\em J. Phys.} {\bf A44} (2011) 015401,
  \href{http://xxx.lanl.gov/abs/1010.1263}{{\tt 1010.1263}}.

\bibitem{Blumenhagen:2011ph}
R.~Blumenhagen, A.~Deser, D.~L{\"u}st, E.~Plauschinn, and F.~Rennecke,
  ``{Non-geometric Fluxes, Asymmetric Strings and Nonassociative Geometry},''
  {\em J. Phys.} {\bf A44} (2011) 385401,
  \href{http://xxx.lanl.gov/abs/1106.0316}{{\tt 1106.0316}}.

\bibitem{Plauschinn:2012kd}
E.~Plauschinn, ``{Non-geometric fluxes and non-associative geometry},'' {\em
  PoS} {\bf CORFU2011} (2011) 061,
  \href{http://xxx.lanl.gov/abs/1203.6203}{{\tt 1203.6203}}.

\bibitem{Mylonas:2012pg}
D.~Mylonas, P.~Schupp, and R.~J. Szabo, ``{Membrane Sigma-Models and
  Quantization of Non-Geometric Flux Backgrounds},'' {\em JHEP} {\bf 09} (2012)
  012, \href{http://xxx.lanl.gov/abs/1207.0926}{{\tt 1207.0926}}.

\bibitem{Bakas:2013jwa}
I.~Bakas and D.~L{\"u}st, ``{3-Cocycles, Non-Associative Star-Products and the
  Magnetic Paradigm of R-Flux String Vacua},'' {\em JHEP} {\bf 01} (2014) 171,
  \href{http://xxx.lanl.gov/abs/1309.3172}{{\tt 1309.3172}}.

\bibitem{Mylonas:2013jha}
D.~Mylonas, P.~Schupp, and R.~J. Szabo, ``{Non-Geometric Fluxes, Quasi-Hopf
  Twist Deformations and Nonassociative Quantum Mechanics},'' {\em J. Math.
  Phys.} {\bf 55} (2014) 122301, \href{http://xxx.lanl.gov/abs/1312.1621}{{\tt
  1312.1621}}.

\bibitem{Chatzistavrakidis:2015vka}
A.~Chatzistavrakidis, L.~Jonke, and O.~Lechtenfeld, ``{Sigma models for
  genuinely non-geometric backgrounds},'' {\em JHEP} {\bf 11} (2015) 182,
  \href{http://xxx.lanl.gov/abs/1505.05457}{{\tt 1505.05457}}.

\bibitem{Szabo:2018hhh}
R.~J. Szabo, ``{Higher Quantum Geometry and Non-Geometric String Theory},'' in
  {\em {17th Hellenic School and Workshops on Elementary Particle Physics and
  Gravity (CORFU2017) Corfu, Greece, September 2-28, 2017}}, 2018.
\newblock \href{http://xxx.lanl.gov/abs/1803.08861}{{\tt 1803.08861}}.

\bibitem{Aldazabal:2006up}
G.~Aldazabal, P.~G. Camara, A.~Font, and L.~E. Ibanez, ``{More dual fluxes and
  moduli fixing},'' {\em JHEP} {\bf 05} (2006) 070,
  \href{http://xxx.lanl.gov/abs/hep-th/0602089}{{\tt hep-th/0602089}}.

\bibitem{Villadoro:2006ia}
G.~Villadoro and F.~Zwirner, ``{D terms from D-branes, gauge invariance and
  moduli stabilization in flux compactifications},'' {\em JHEP} {\bf 03} (2006)
  087, \href{http://xxx.lanl.gov/abs/hep-th/0602120}{{\tt hep-th/0602120}}.

\bibitem{Micu:2007rd}
A.~Micu, E.~Palti, and G.~Tasinato, ``{Towards Minkowski Vacua in Type II
  String Compactifications},'' {\em JHEP} {\bf 03} (2007) 104,
  \href{http://xxx.lanl.gov/abs/hep-th/0701173}{{\tt hep-th/0701173}}.

\bibitem{Font:2008vd}
A.~Font, A.~Guarino, and J.~M. Moreno, ``{Algebras and non-geometric flux
  vacua},'' {\em JHEP} {\bf 12} (2008) 050,
  \href{http://xxx.lanl.gov/abs/0809.3748}{{\tt 0809.3748}}.

\bibitem{Caviezel:2009tu}
C.~Caviezel, T.~Wrase, and M.~Zagermann, ``{Moduli Stabilization and Cosmology
  of Type IIB on SU(2)-Structure Orientifolds},'' {\em JHEP} {\bf 04} (2010)
  011, \href{http://xxx.lanl.gov/abs/0912.3287}{{\tt 0912.3287}}.

\bibitem{Dibitetto:2011gm}
G.~Dibitetto, A.~Guarino, and D.~Roest, ``{Charting the landscape of N=4 flux
  compactifications},'' {\em JHEP} {\bf 03} (2011) 137,
  \href{http://xxx.lanl.gov/abs/1102.0239}{{\tt 1102.0239}}.

\bibitem{Hassler:2014mla}
F.~Hassler, D.~L{\"u}st, and S.~Massai, ``{On Inflation and de Sitter in
  Non-Geometric String Backgrounds},'' {\em Fortsch. Phys.} {\bf 65} (2017),
  no.~10-11 1700062, \href{http://xxx.lanl.gov/abs/1405.2325}{{\tt 1405.2325}}.

\bibitem{Blumenhagen:2015kja}
R.~Blumenhagen, A.~Font, M.~Fuchs, D.~Herschmann, E.~Plauschinn, Y.~Sekiguchi,
  and F.~Wolf, ``{A Flux-Scaling Scenario for High-Scale Moduli Stabilization
  in String Theory},'' {\em Nucl. Phys.} {\bf B897} (2015) 500--554,
  \href{http://xxx.lanl.gov/abs/1503.07634}{{\tt 1503.07634}}.

\bibitem{Lawrence:2006ma}
A.~Lawrence, M.~B. Schulz, and B.~Wecht, ``{D-branes in nongeometric
  backgrounds},'' {\em JHEP} {\bf 07} (2006) 038,
  \href{http://xxx.lanl.gov/abs/hep-th/0602025}{{\tt hep-th/0602025}}.

\bibitem{Albertsson:2008gq}
C.~Albertsson, T.~Kimura, and R.~A. Reid-Edwards, ``{D-branes and doubled
  geometry},'' {\em JHEP} {\bf 04} (2009) 113,
  \href{http://xxx.lanl.gov/abs/0806.1783}{{\tt 0806.1783}}.

\bibitem{Buscher:1987sk}
T.~Buscher, ``{A Symmetry of the String Background Field Equations},'' {\em
  Phys.Lett.} {\bf B194} (1987) 59.

\bibitem{Alvarez:1996up}
E.~Alvarez, J.~L.~F. Barbon, and J.~Borlaf, ``{T duality for open strings},''
  {\em Nucl. Phys.} {\bf B479} (1996) 218--242,
  \href{http://xxx.lanl.gov/abs/hep-th/9603089}{{\tt hep-th/9603089}}.

\bibitem{Dorn:1996an}
H.~Dorn and H.~J. Otto, ``{On T duality for open strings in general Abelian and
  nonAbelian gauge field backgrounds},'' {\em Phys. Lett.} {\bf B381} (1996)
  81--88, \href{http://xxx.lanl.gov/abs/hep-th/9603186}{{\tt hep-th/9603186}}.

\bibitem{Dorn:1997if}
H.~Dorn and H.~J. Otto, ``{Remarks on T duality for open strings},'' {\em Nucl.
  Phys. Proc. Suppl.} {\bf 56B} (1997) 30--35,
  \href{http://xxx.lanl.gov/abs/hep-th/9702018}{{\tt hep-th/9702018}}.

\bibitem{Forste:1996hy}
S.~F{\"o}rste, A.~Kehagias, and S.~Schwager, ``Non-abelian duality for open
  strings,'' {\em Nuclear Physics B} {\bf 478} (1996), no.~1 141 -- 155.

\bibitem{Forste:1996ai}
S.~F{\"o}rste, A.~A. Kehagias, and S.~Schwager, ``{NonAbelian T duality for
  open strings},'' {\em Nucl. Phys. Proc. Suppl.} {\bf 56B} (1997) 36--41,
  \href{http://xxx.lanl.gov/abs/hep-th/9610062}{{\tt hep-th/9610062}}.

\bibitem{Forste:1996ms}
S.~F{\"o}rste, A.~A. Kehagias, and S.~Schwager, ``{T duality for open strings
  with respect to nonAbelian isometries},'' in {\em {Gauge theories, applied
  supersymmetry and quantum gravity. Proceedings, 2nd Conference, London, UK,
  July 5-10, 1996}}, pp.~271--278, 1996.
\newblock \href{http://xxx.lanl.gov/abs/hep-th/9611060}{{\tt hep-th/9611060}}.

\bibitem{Albertsson:2004gr}
C.~Albertsson, U.~Lindstrom, and M.~Zabzine, ``{T-duality for the sigma model
  with boundaries},'' {\em JHEP} {\bf 12} (2004) 056,
  \href{http://xxx.lanl.gov/abs/hep-th/0410217}{{\tt hep-th/0410217}}.

\bibitem{Borlaf:1996na}
J.~Borlaf and Y.~Lozano, ``{Aspects of T duality in open strings},'' {\em Nucl.
  Phys.} {\bf B480} (1996) 239--264,
  \href{http://xxx.lanl.gov/abs/hep-th/9607051}{{\tt hep-th/9607051}}.

\bibitem{Lozano:1996sc}
Y.~Lozano, ``{Duality and canonical transformations},'' {\em Mod. Phys. Lett.}
  {\bf A11} (1996) 2893--2914,
  \href{http://xxx.lanl.gov/abs/hep-th/9610024}{{\tt hep-th/9610024}}.

\bibitem{Tseytlin:1996it}
A.~A. Tseytlin, ``{Selfduality of Born-Infeld action and Dirichlet three-brane
  of type IIB superstring theory},'' {\em Nucl. Phys.} {\bf B469} (1996)
  51--67, \href{http://xxx.lanl.gov/abs/hep-th/9602064}{{\tt hep-th/9602064}}.

\bibitem{Bergshoeff:1996cy}
E.~Bergshoeff and M.~De~Roo, ``{D-branes and T duality},'' {\em Phys. Lett.}
  {\bf B380} (1996) 265--272,
  \href{http://xxx.lanl.gov/abs/hep-th/9603123}{{\tt hep-th/9603123}}.

\bibitem{Green:1996bh}
M.~B. Green, C.~M. Hull, and P.~K. Townsend, ``{D-brane Wess-Zumino actions, t
  duality and the cosmological constant},'' {\em Phys. Lett.} {\bf B382} (1996)
  65--72, \href{http://xxx.lanl.gov/abs/hep-th/9604119}{{\tt hep-th/9604119}}.

\bibitem{Kawai:2007qd}
S.~Kawai and Y.~Sugawara, ``{D-branes in T-fold conformal field theory},'' {\em
  JHEP} {\bf 02} (2008) 027, \href{http://xxx.lanl.gov/abs/0709.0257}{{\tt
  0709.0257}}.

\bibitem{Grange:2005nm}
P.~Grange and R.~Minasian, ``{Tachyon condensation and D-branes in generalized
  geometries},'' {\em Nucl. Phys.} {\bf B741} (2006) 199--214,
  \href{http://xxx.lanl.gov/abs/hep-th/0512185}{{\tt hep-th/0512185}}.

\bibitem{Klimcik:1995np}
C.~Klimcik and P.~Severa, ``{Poisson Lie T duality: Open strings and
  D-branes},'' {\em Phys. Lett.} {\bf B376} (1996) 82--89,
  \href{http://xxx.lanl.gov/abs/hep-th/9512124}{{\tt hep-th/9512124}}.

\bibitem{Albertsson:2006zg}
C.~Albertsson and R.~A. Reid-Edwards, ``{Worldsheet boundary conditions in
  Poisson-Lie T-duality},'' {\em JHEP} {\bf 03} (2007) 004,
  \href{http://xxx.lanl.gov/abs/hep-th/0606024}{{\tt hep-th/0606024}}.

\bibitem{Davidovic:2016xhh}
L.~Davidovic, ``{Open string T-duality in a weakly curved background},'' {\em
  Eur. Phys. J.} {\bf C76} (2016), no.~12 660,
  \href{http://xxx.lanl.gov/abs/1603.06411}{{\tt 1603.06411}}.

\bibitem{Sazdovic:2016ley}
B.~Sazdovic, ``{From geometry to non-geometry via T-duality},''
  \href{http://xxx.lanl.gov/abs/1606.01938}{{\tt 1606.01938}}.

\bibitem{Sazdovic:2017lqo}
B.~Sazdovic, ``{Open string T-duality in double space},'' {\em Eur. Phys. J.}
  {\bf C77} (2017), no.~9 634, \href{http://xxx.lanl.gov/abs/1704.01163}{{\tt
  1704.01163}}.

\bibitem{Cappell2006}
S.~{Cappell}, D.~{DeTurck}, H.~{Gluck}, and E.~Y. {Miller}, ``{Cohomology of
  Harmonic Forms on Riemannian Manifolds With Boundary},'' {\em ArXiv
  Mathematics e-prints} (Aug., 2005)
  \href{http://xxx.lanl.gov/abs/math/0508372}{{\tt math/0508372}}.

\bibitem{Rocek:1991ps}
M.~Rocek and E.~P. Verlinde, ``{Duality, quotients, and currents},'' {\em Nucl.
  Phys.} {\bf B373} (1992) 630--646,
  \href{http://xxx.lanl.gov/abs/hep-th/9110053}{{\tt hep-th/9110053}}.

\bibitem{Giveon:1993ai}
A.~Giveon and M.~Rocek, ``{On nonAbelian duality},'' {\em Nucl. Phys.} {\bf
  B421} (1994) 173--190, \href{http://xxx.lanl.gov/abs/hep-th/9308154}{{\tt
  hep-th/9308154}}.

\bibitem{Alvarez:1993qi}
E.~Alvarez, L.~Alvarez-Gaume, J.~L.~F. Barbon, and Y.~Lozano, ``{Some global
  aspects of duality in string theory},'' {\em Nucl. Phys.} {\bf B415} (1994)
  71--100, \href{http://xxx.lanl.gov/abs/hep-th/9309039}{{\tt hep-th/9309039}}.

\bibitem{Plauschinn:2014nha}
E.~Plauschinn, ``{On T-duality transformations for the three-sphere},'' {\em
  Nucl. Phys.} {\bf B893} (2015) 257--286,
  \href{http://xxx.lanl.gov/abs/1408.1715}{{\tt 1408.1715}}.

\bibitem{Buscher:1987qj}
T.~Buscher, ``{Path Integral Derivation of Quantum Duality in Nonlinear Sigma
  Models},'' {\em Phys.Lett.} {\bf B201} (1988) 466.

\bibitem{Plauschinn:2013wta}
E.~Plauschinn, ``{T-duality revisited},'' {\em JHEP} {\bf 01} (2014) 131,
  \href{http://xxx.lanl.gov/abs/1310.4194}{{\tt 1310.4194}}.

\bibitem{Freed:1999vc}
D.~S. Freed and E.~Witten, ``{Anomalies in string theory with D-branes},'' {\em
  Asian J. Math.} {\bf 3} (1999) 819,
  \href{http://xxx.lanl.gov/abs/hep-th/9907189}{{\tt hep-th/9907189}}.

\bibitem{Camara:2005dc}
P.~G. Camara, A.~Font, and L.~E. Ibanez, ``{Fluxes, moduli fixing and MSSM-like
  vacua in a simple IIA orientifold},'' {\em JHEP} {\bf 09} (2005) 013,
  \href{http://xxx.lanl.gov/abs/hep-th/0506066}{{\tt hep-th/0506066}}.

\bibitem{LoaizaBrito:2006se}
O.~Loaiza-Brito, ``{Freed-witten anomaly in general flux compactification},''
  {\em Phys. Rev.} {\bf D76} (2007) 106015,
  \href{http://xxx.lanl.gov/abs/hep-th/0612088}{{\tt hep-th/0612088}}.

\bibitem{Aldazabal:2008zza}
G.~Aldazabal, P.~G. Camara, and J.~A. Rosabal, ``{Flux algebra, Bianchi
  identities and Freed-Witten anomalies in F-theory compactifications},'' {\em
  Nucl. Phys.} {\bf B814} (2009) 21--52,
  \href{http://xxx.lanl.gov/abs/0811.2900}{{\tt 0811.2900}}.

\bibitem{Aldazabal:2011yz}
G.~Aldazabal, D.~Marques, C.~Nunez, and J.~A. Rosabal, ``{On Type IIB moduli
  stabilization and N = 4, 8 supergravities},'' {\em Nucl. Phys.} {\bf B849}
  (2011) 80--111, \href{http://xxx.lanl.gov/abs/1101.5954}{{\tt 1101.5954}}.

\bibitem{delaOssa:1992vci}
X.~C. de~la Ossa and F.~Quevedo, ``{Duality symmetries from nonAbelian
  isometries in string theory},'' {\em Nucl. Phys.} {\bf B403} (1993) 377--394,
  \href{http://xxx.lanl.gov/abs/hep-th/9210021}{{\tt hep-th/9210021}}.

\bibitem{Sfetsos:1994vz}
K.~Sfetsos, ``{Gauged WZW models and nonAbelian duality},'' {\em Phys. Rev.}
  {\bf D50} (1994) 2784--2798,
  \href{http://xxx.lanl.gov/abs/hep-th/9402031}{{\tt hep-th/9402031}}.

\bibitem{Alvarez:1994np}
E.~Alvarez, L.~Alvarez-Gaume, and Y.~Lozano, ``{On nonAbelian duality},'' {\em
  Nucl. Phys.} {\bf B424} (1994) 155--183,
  \href{http://xxx.lanl.gov/abs/hep-th/9403155}{{\tt hep-th/9403155}}.

\bibitem{Lozano:1995jx}
Y.~Lozano, ``{NonAbelian duality and canonical transformations},'' {\em Phys.
  Lett.} {\bf B355} (1995) 165--170,
  \href{http://xxx.lanl.gov/abs/hep-th/9503045}{{\tt hep-th/9503045}}.

\end{thebibliography}\endgroup
\bibliographystyle{utphys}


\end{document}